\documentstyle[12pt,epsfig]{article}

\begin{document}

\newcommand{\lsim}{\raisebox{-0.13cm}{~\shortstack{$<$ \\[-0.07cm] $\sim$}}~}
\newcommand{\gsim}{\raisebox{-0.13cm}{~\shortstack{$>$ \\[-0.07cm] $\sim$}}~}
\newcommand{\nn}{\noindent}
\newcommand{\non}{\nonumber}
\newcommand{\ee}{e^+ e^-}
\newcommand{\ra}{\rightarrow}
\newcommand{\tb}{\tan \beta}
\newcommand{\s}{\smallskip}
\newcommand{\beq}{\begin{eqnarray}}
\newcommand{\eeq}{\end{eqnarray}}

\vspace*{.1cm} 
\baselineskip=17pt

\begin{flushright}
PM/02--13\\
May 2002\\
\end{flushright}

\vspace*{0.9cm}

\begin{center}

{\large\sc {\bf The Search for Higgs particles at high--energy colliders:}}

\vspace*{2mm}

{\large\sc {\bf Past, Present and Future}}

\vspace{0.7cm}

{\sc Abdelhak DJOUADI} 
\vspace*{5mm} 

Laboratoire de Physique Math\'ematique et Th\'eorique, UMR5825--CNRS,\\
Universit\'e de Montpellier II, F--34095 Montpellier Cedex 5, France. 
\end{center} 

\vspace*{1cm} 

\begin{abstract}
\nn I briefly review the Higgs sector in the Standard Model and its minimal 
Supersymmetric extension, the MSSM. After summarizing the properties of the 
Higgs bosons and the present experimental constraints, I will discuss the 
prospects for discovering these particle at the upgraded Tevatron, the LHC 
and a high--energy $e^+e^-$ linear collider. The possibility of studying the
properties of the Higgs particles will be then summarized. 
\end{abstract}
\vspace*{1.8cm}

\begin{center}
{\it Review talk given at the Workshop on High--Energy Physics Phenomenology,} 

{\it WHEPP VII,  Harish-Chandra Research Institute,}

{\it Allahabad, India, January 4--15, 2002}
\end{center}

\newpage

\subsection*{1. Introduction}

The search for Higgs bosons is one of the main missions of present and
future high--energy colliders. The observation of this particle is of major
importance for the present understanding of the interactions of the fundamental
particles. Indeed, in order to accommodate the well--established electromagnetic
and weak interaction phenomena, the existence of at least one isodoublet scalar
field to generate fermion and weak gauge bosons masses is required. The
Standard Model (SM) makes use of one isodoublet field: three Goldstone bosons
among the four degrees of freedom are absorbed to build up the longitudinal
components of the massive $W^\pm,Z$ gauge bosons; one degree of freedom is left
over corresponding to a physical scalar particle, the Higgs boson \cite{R0}. 
Despite of its numerous successes in explaining the present data, the SM will
not be completely tested before this particle has been experimentally observed
and its fundamental properties studied.  

In the SM, the profile of the Higgs particle is uniquely determined once its
mass $M_H$ is fixed \cite{R1}. The decay width, the branching ratios and the
production cross sections are given by the strength of the Yukawa couplings to
fermions and gauge bosons, the scale of which is set by the masses of these
particles. Unfortunately, the Higgs boson is a free parameter.  

The only available information on $M_H$ is the upper limit $M_H \geq 114.1$ GeV
established at LEP2 \cite{R2}. [The collaborations have also reported a
2.1$\sigma$ excess of events beyond the expected SM backgrounds consistent with
a SM--like Higgs boson with a mass $M_H \sim 115$ GeV \cite{R2}, as will be
discussed later.] Furthermore, the accuracy of the electroweak data measured at
LEP, SLC and the Tevatron provides sensitivity to $M_H$: the Higgs boson
contributes logarithmically, $\propto \log (M_H/M_W)$, to the radiative
corrections to the $W/Z$ boson propagators. A recent analysis yields the value
$M_H=88^{+60}_{-37}$ GeV, corresponding to a 95\% CL upper limit of $M_H \lsim
206$ GeV \cite{R3} [there is still an error due to the hadronic contribution
to the running of the fine structure constant $\alpha$]; see the left panel in
Fig.~1.  

However, interesting theoretical constraints can be derived from assumptions on
the energy range within which the Standard Model is valid before perturbation 
theory breaks down and new phenomena would emerge.  

$(i)$ If the Higgs mass were larger than $\sim$ 1 TeV, the $W$ and $Z$ 
bosons would interact strongly with each other to ensure unitarity 
in their scattering at high energies. Imposing the unitarity requirement 
in the scattering of longitudinal $W$ bosons at high--energy for instance 
leads to the tree--level bound $M_H \lsim 870$ GeV \cite{R4a}. Note also 
that radiative corrections to the Higgs boson couplings become non-perturbative 
for masses beyond $M_H \gsim 1$ TeV \cite{R4} and the Higgs resonance becomes
too wide as will be discussed later. 

$(ii)$ The quartic Higgs self--coupling, which at the scale $M_H$ is fixed by
$M_H$ itself, grows logarithmically with the energy scale. If $M_H$ is small,
the energy cut--off $\Lambda$ at which the coupling grows beyond any bound and
new phenomena should occur, is large; if $M_H$ is large, $\Lambda$ is small. 
The condition $M_H \lsim \Lambda$ sets an upper limit on the Higgs mass in the
SM; lattice analyses lead to an estimate of about $M_H \sim 630$ GeV for this
limit.  Furthermore, top quark loops tend to drive the coupling to negative
values for which the vacuum is no more stable. Therefore, requiring the SM to
be extended to the GUT scale, $\Lambda_{\rm GUT} \sim 10^{16}$ GeV, and
including the effect of top quark loops on the running coupling, the Higgs mass
should roughly lie in the range 130 GeV $\lsim M_H \lsim 180$ GeV \cite{R4};
see right panel of Fig.~1.

\begin{figure}[hb!]
\vspace*{-.7cm}
\begin{center}
\begin{tabular}{c c}
\hspace*{-7mm}
\epsfig{file=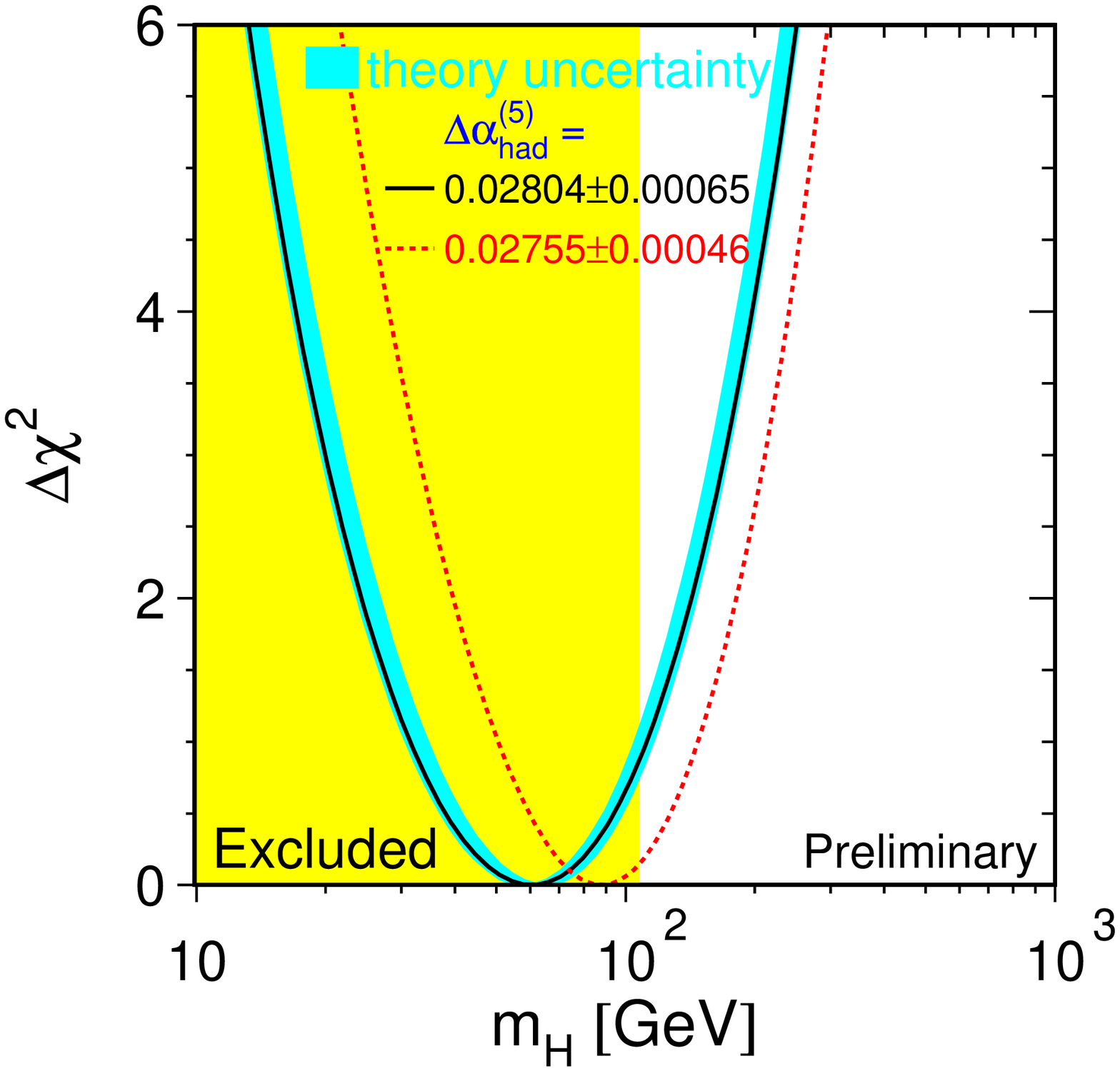,width=0.5\linewidth,height=10.3cm} &
\epsfig{file=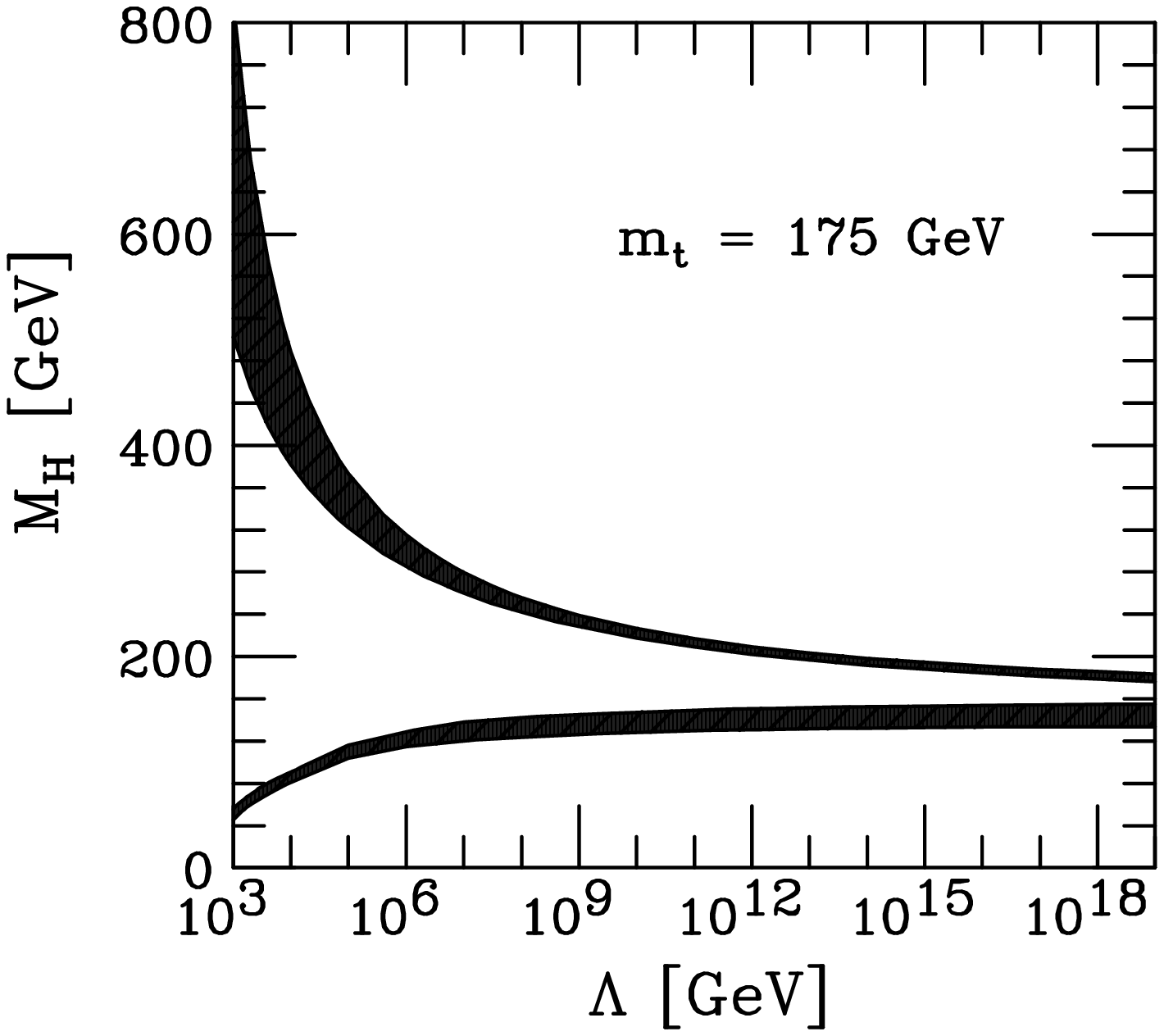,width=0.5\linewidth,height=9.9cm} \\
\end{tabular}
\vspace{-0.8cm}
\caption{The $\chi^2$ of the fit to electroweak data as a function of $M_H$ 
(left) and triviality and vacuum stability bounds on $M_H$ as a function of 
the new physics scale $\Lambda$ (right). } 
\end{center}
\vspace*{-.3cm}
\label{fig1}
\end{figure} 

However, there are two problems that one has to face when trying to extend the
SM to $\Lambda_{\rm GUT}$. The first one is the so--called hierarchy or
naturalness problem: the Higgs boson tends to acquire a mass of the order of
these large scales [the radiative corrections to $M_H$ are quadratically
divergent]; the second problem is that the simplest GUTs predict a value for
$\sin^2\theta_W$ that is incompatible with the measured value $\sin^2\theta_W
\simeq~0.23$. Low energy supersymmetry solves these two problems at once:
supersymmetric particle loops cancel exactly the quadratic divergences and
contribute to the running of the gauge coupling constants, correcting the small
discrepancy to the observed value of $\sin^2\theta_W$ \cite{MSSM}. \bigskip

The Minimal Supersymmetric extension of the Standard Model (MSSM) \cite{MSSM}
requires the existence of two isodoublets of Higgs fields, to cancel anomalies
and to give mass separately to up and down--type fermions. Two CP--even neutral
Higgs bosons $h,H$, a pseudoscalar $A$ bosons and a pair of charged scalar
particles, $H^\pm$, are introduced by this extension of the Higgs sector
\cite{R1}. Besides the four masses, two additional parameters define the
properties of these particles: a mixing angle $\alpha$ in the neutral CP--even
sector and the ratio of the two vacuum expectation values $\tb$, which from GUT
restrictions is assumed in the range $1 \lsim \tb \lsim m_t/m_b$ with the lower
and upper ranges favored by Yukawa coupling unification [the lower range is
excluded by LEP2 searches].  

Supersymmetry leads to several relations among these parameters and only two of
them, taken in general as $M_A$ and $\tb$ are in fact independent. These
relations impose a strong hierarchical structure on the mass spectrum,
$M_h<M_Z, M_A<M_H$ and $M_W<M_{H^\pm}$, which however is broken by radiative
corrections if the top quark mass is large \cite{R5}. The leading part of this
correction grows as the fourth power of $m_t$ and logarithmically with the
squark mass $M_S$; the mixing (or trilinear coupling) in the stop sector $A_t$
plays an important role. For instance, the upper bound on the mass of the
lightest Higgs boson $h$ is shifted from the tree level value $M_Z$ to $M_h
\sim 130$ GeV for $A_t=\sqrt{6}M_S$ with $M_S=1$ TeV \cite{R5}. The masses of
the heavy neutral and charged Higgs particles are expected to be in the range
of the electroweak symmetry breaking scale; see left panel of Fig.~2.

\begin{figure}[htbp]
\begin{center}
\vspace*{-.8cm}
\epsfig{file=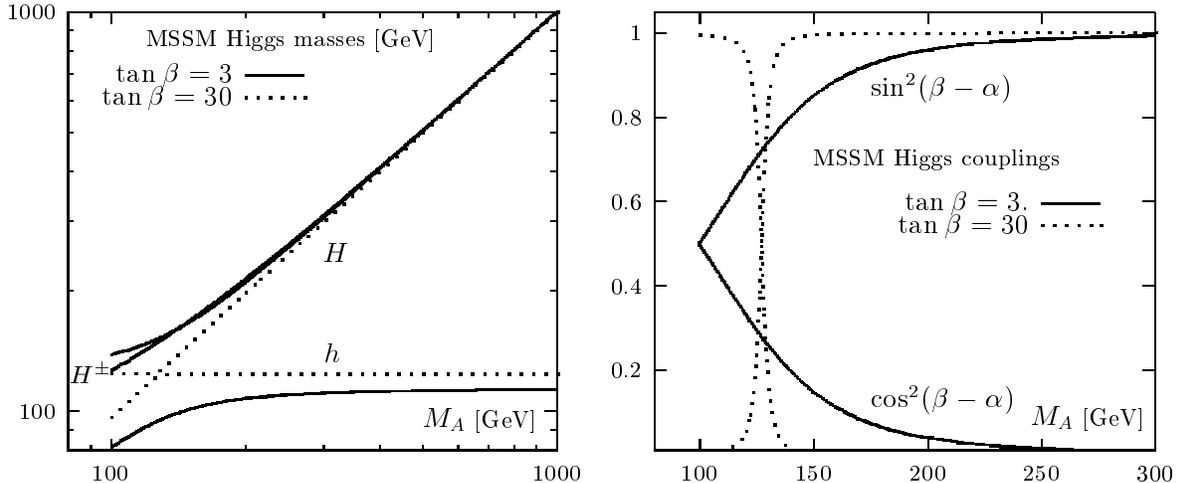,bbllx=80,bblly=470,bburx=570,bbury=700,height=7.cm,width=16cm}
\caption{The masses of the Higgs bosons in the MSSM and their relative squared 
couplings to the massive gauge bosons for two representative values $\tan 
\beta$=3 and 30.}
\end{center}
\vspace*{-.5cm}
\label{fig2}
\end{figure}      
The couplings of the various neutral Higgs bosons [collectively denoted
by $\Phi$] to fermions and gauge bosons will in general strongly depend
on the angles $\alpha$ and $\beta$; normalized to the SM Higgs
boson couplings, they are given by:
\begin{center}
\begin{tabular}{|c|c|c|c|c|} \hline
$\ \ \ \Phi \ \ \ $ &$ g_{\Phi \bar{u}u} $	& $ g_{\Phi \bar{d} d} $ &
$g_{ \Phi VV} $ \\ \hline
$h$  & \ $\; \cos\alpha/\sin\beta	\ \to 1 $ \ & \ $ \;	-\sin\alpha/
\cos\beta \ \to 1 $ \ & \ $ \; \sin(\beta-\alpha) \ \to  1	$ \ \\
 $H$  & \	$\; \sin\alpha/\sin\beta \ \to 1/\tb $ \ & \ $ \; \cos\alpha/
\cos\beta \ \to \tb $ \ & \ $ \; \cos(\beta-\alpha) \ \to 0	$ \ \\
$A$  & \ $\; 1/ \tb \; $\ & \ $	\; \tb \; $ \	& \ $ \; 0 \; $	\ \\ \hline
\end{tabular}
\end{center}
\vspace*{2mm}

The pseudoscalar has no tree level couplings to gauge bosons, and its
couplings to down (up) type fermions are (inversely) proportional to
$\tb$. It is also the case for the couplings of the charged Higgs
particle to fermions which are a mixture of scalar and pseudoscalar
currents and depend only on $\tb$. For the CP--even Higgs bosons, the
couplings to down (up) type fermions are enhanced (suppressed) compared
to the SM Higgs couplings for $\tb >1$. They share the SM Higgs 
couplings to vector bosons since they are suppressed by $\sin (\beta-\alpha)$
and $\cos(\beta-\alpha)$ factors, respectively for $h$ and $H$; see 
right panel of Fig.~2. 

If the pseudoscalar mass is large, the $h$ boson mass reaches its upper limit
[which depends on the value of $\tb$] and its couplings to fermions and gauge
bosons are SM like; the heavier CP--even $H$ and charged $H^\pm$ bosons become
degenerate with $A$. In this decoupling limit, it is very difficult to
distinguish the Higgs sectors of the SM and MSSM. 

Let us summarze the constraints on the MSSM Higgs particles masses, which
mainly come from the negative LEP2 searches \cite{R2} in the Higgs--strahlung,
$\ee \to Z+h/H$, and pair production, $\ee \to A+h/H$, processes which will be
discussed in more detail later. In the decoupling limit where the $h$ boson has
SM--like couplings to $Z$ bosons, the limit $M_h \gsim 114.1$ GeV from the $\ee
\to hZ$ process holds. This constraint rules out $\tb$ values larger than $\tb
\gsim 3$. From the $\ee \to Ah$ process, one obtains the absolute limits $M_h
\gsim 91$ GeV and $M_A \gsim 92$ GeV, for a maximal $ZhA$ coupling. In the
general case, the allowed values for $M_h$ are shown in Fig.~3 as a function of
$\tan \beta$ (the colored regions) in the cases of maximal, typical and
no--mixing in the stop sector, respectively $A_t=\sqrt{6}M_S, M_S$ and $0$
for $M_S=1$ TeV. Also are shown, the implications  of the $2.1 \sigma$ evidence
for a SM--like Higgs boson with a mass $115.6^{+1.3}_{-0.9}$ GeV. 
Allowing for an error on the Higgs mass and an almost maximal coupling to the
$Z$ boson, the red (green) region indicates where $114$~GeV$<M_{h}(M_{H}) 
<117$~GeV and $\sin(\cos)^2(\beta-\alpha)>0.9$. 

\begin{figure}[htbp]
\vspace*{-5mm}
\begin{center}
\epsfig{figure=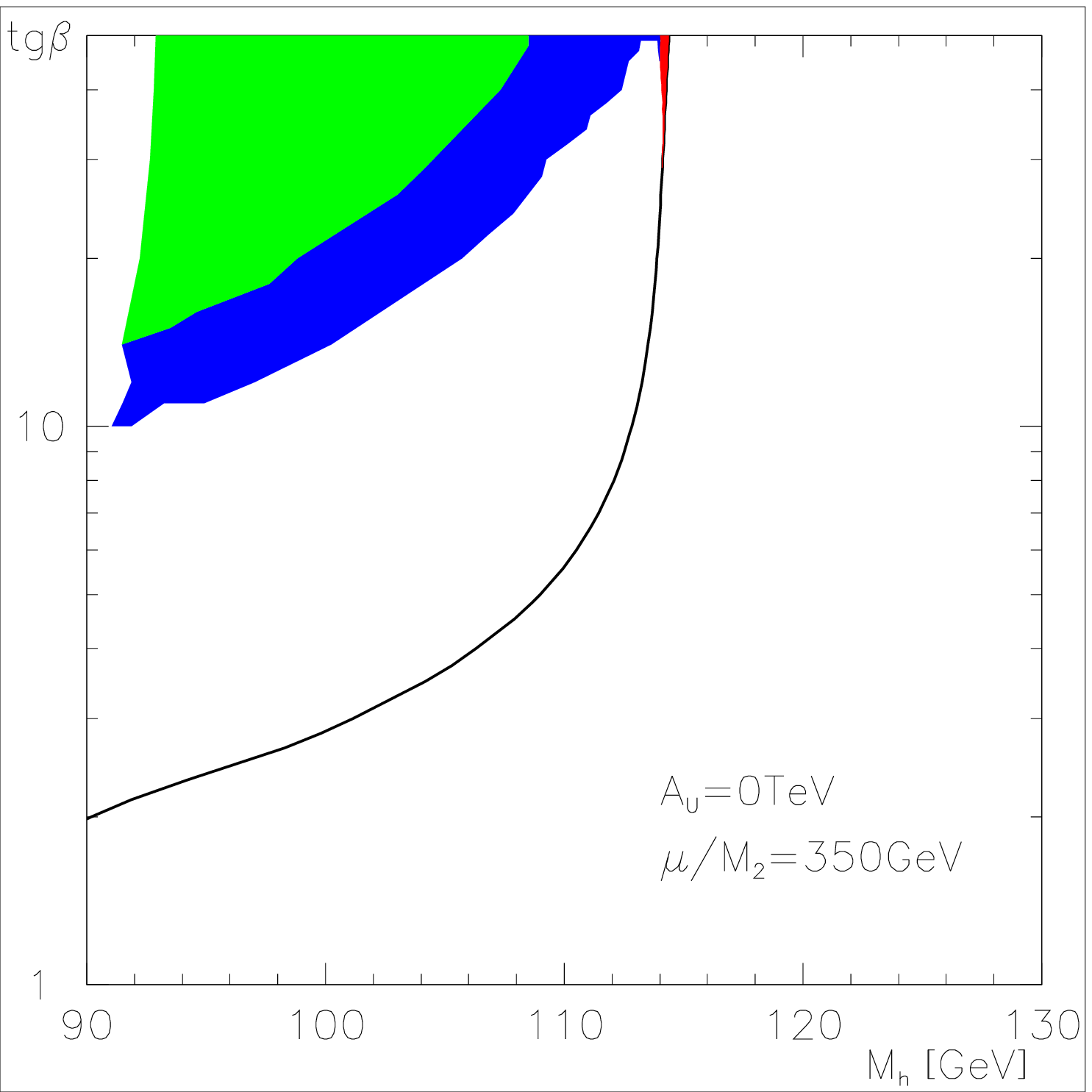,bbllx=3,bblly=3,bburx=410,bbury=420,height=5.9cm,width=5.1cm,clip=}
%\hspace{.5cm}
\epsfig{figure=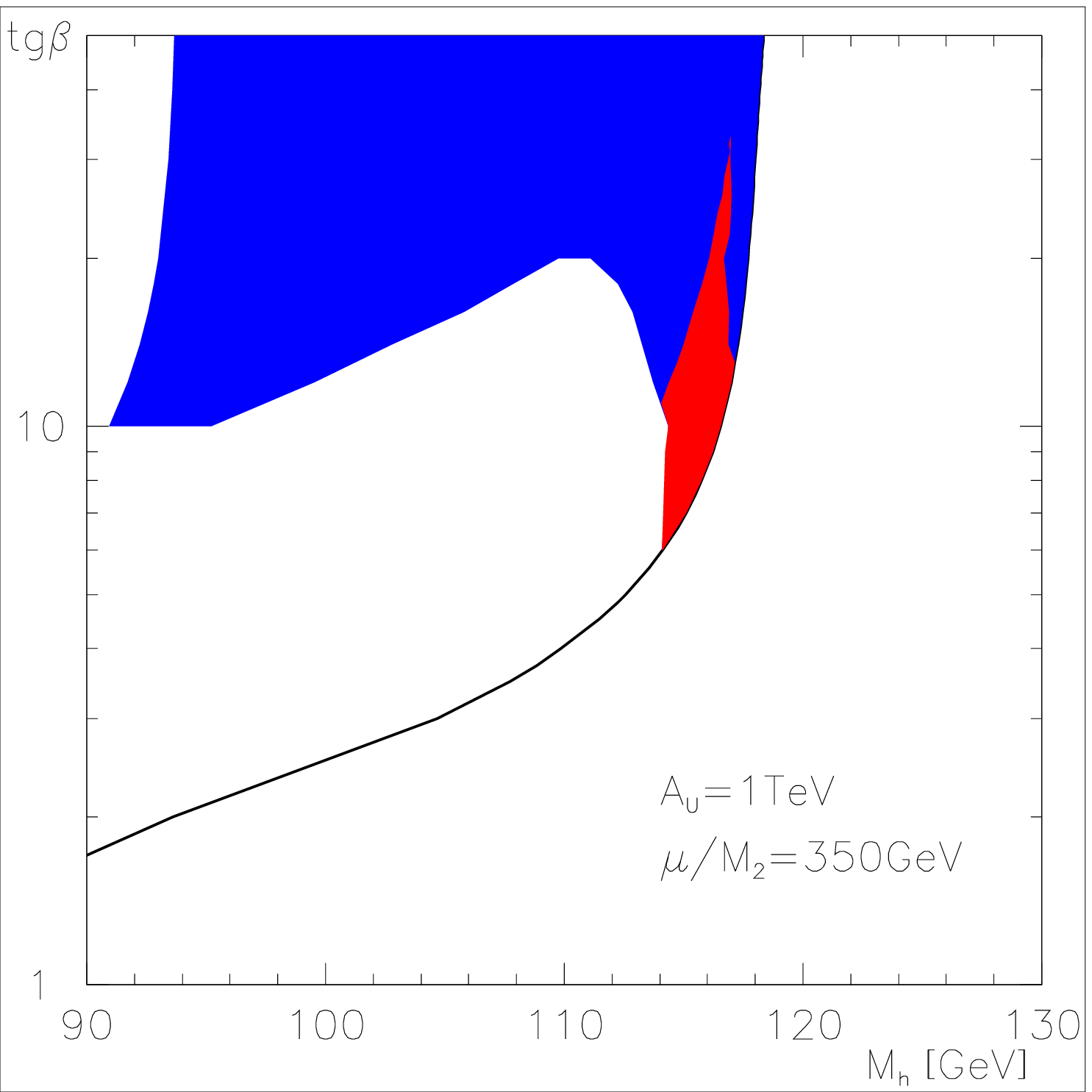,bbllx=3,bblly=3,bburx=410,bbury=420,height=5.9cm,width=5.1cm,clip=}
%\hspace{.5cm}
\epsfig{figure=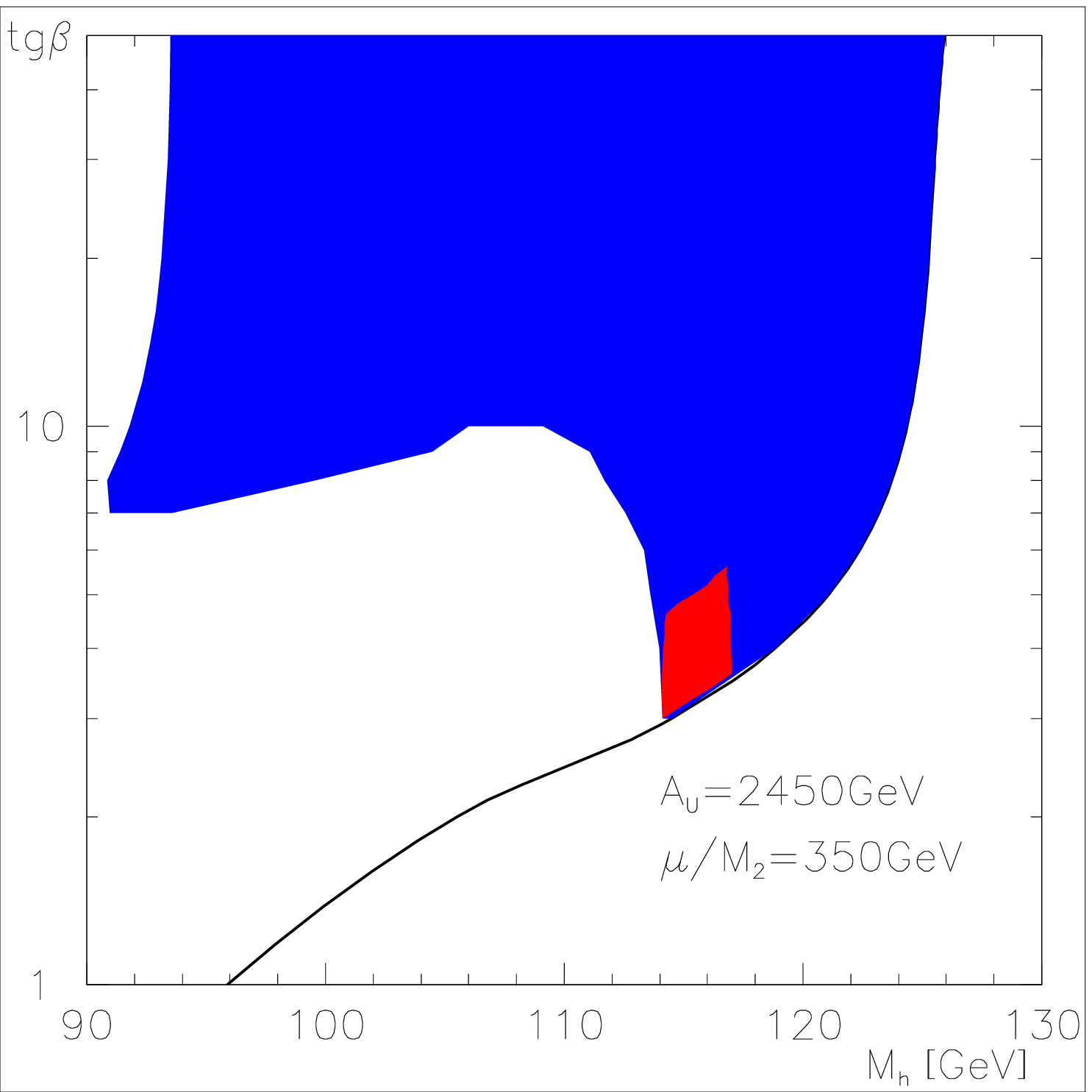,bbllx=3,bblly=3,bburx=410,bbury=420,height=5.9cm,width=5.1cm,clip=}
\vspace*{1mm}
\caption{The allowed (colored) regions for $M_h$ from LEP2 searches as a 
function of $\tan \beta$ in the case of maximal, typical and no stop mixing;
From Ref.~[7].} 
\end{center}
\vspace*{-.4cm}
\end{figure}

In more general SUSY scenarii, one can add an arbitrary number of Higgs 
doublet and/or singlet fields without being in conflict with high precision 
data \cite{R3}. The Higgs spectrum becomes then much more complicated than 
in the MSSM, and much less constrained. However, the triviality argument
always imposes a bound on the mass of the lightest Higgs boson of the theory. 
For instance, if only one Higgs singlet field is added to the MSSM, an upper
bound $M_h \lsim 150$ GeV can be set \cite{R6a}. In the most general
SUSY model, with arbitrary matter content and gauge coupling unification 
near $\Lambda_{\rm GUT}$, and absolute upper limit on the mass of the lightest 
Higgs boson, $M_h \lsim 205$ GeV, has been derived \cite{R6}.

Thus, either in the SM or in its SUSY extensions, a Higgs boson should be
lighter than $\sim 200$ GeV, and will be therefore kinematically accessible at
the next generation of experiments.  In the following, after summarizing the
decay modes of the Higgs bosons, I will briefly discuss the discovery potential
of present and future colliders, the Tevatron Run II \cite{R8}, the LHC
\cite{R9,R10} and a future ${\rm \ee}$ linear collider \cite{R11} with a c.m. 
energy in the range of 300 to 800 GeV such as the TESLA machine \cite{R12}.

\subsection*{2. Decay Modes}

Let us first discuss the Higgs decay modes relying on the analyses of
Ref.~\cite{R13}; see Fig.~4. To simplify the discussion in the SM, it is
convenient  to divide the Higgs mass into two ranges: the ``low mass" range
$M_H \lsim 130$ GeV and the ``high mass" range $M_H \gsim 130$ GeV.  

In the ``low mass" range, the Higgs boson decays into a large variety of
channels. The main decay mode is by far the decay into $b\bar{b}$ pairs
with a branching ratio of $\sim 90\%$ followed by the decays into
$c\bar{c}$ and $\tau^+\tau^-$ pairs with branching ratios of $\sim
5\%$. Also of significance, the top--loop mediated Higgs decay into
gluons, which for $M_H$ around 120 GeV occurs at the level of $\sim
5\%$. The top and $W$--loop mediated $\gamma\gamma$ and $Z \gamma$ decay
modes are very rare the branching ratios being of ${\cal O }(10^{-3})$.
However, these decays lead to clear signals and are theoretically interesting 
being sensitive to new heavy particles such as SUSY particles. 

In the ``high mass" range, the Higgs bosons decay into $WW$ and $ZZ$
pairs, with one of the gauge bosons being virtual below the threshold.
Above the $ZZ$ threshold, the Higgs boson decays almost exclusively into
these channels with a branching ratio of 2/3 for $WW$ and 1/3 for $ZZ$.
The opening of the $t\bar{t}$ channel does not alter significantly this
pattern. 

In the low mass range, the Higgs boson is very narrow $\Gamma_H<10$ MeV,
but the width becomes rapidly wider for masses larger than 130 GeV,
reaching 1 GeV at the $ZZ$ threshold. The Higgs total width cannot be
measured directly in the mass range below 250 GeV. For large masses, 
$M_H \gsim 500$ GeV, the Higgs boson becomes obese since its total
width is comparable to its mass, and it is hard to consider the Higgs 
as a resonance. 

\begin{figure}[htbp]
\begin{center}
\vspace*{-.5cm}
\epsfig{file=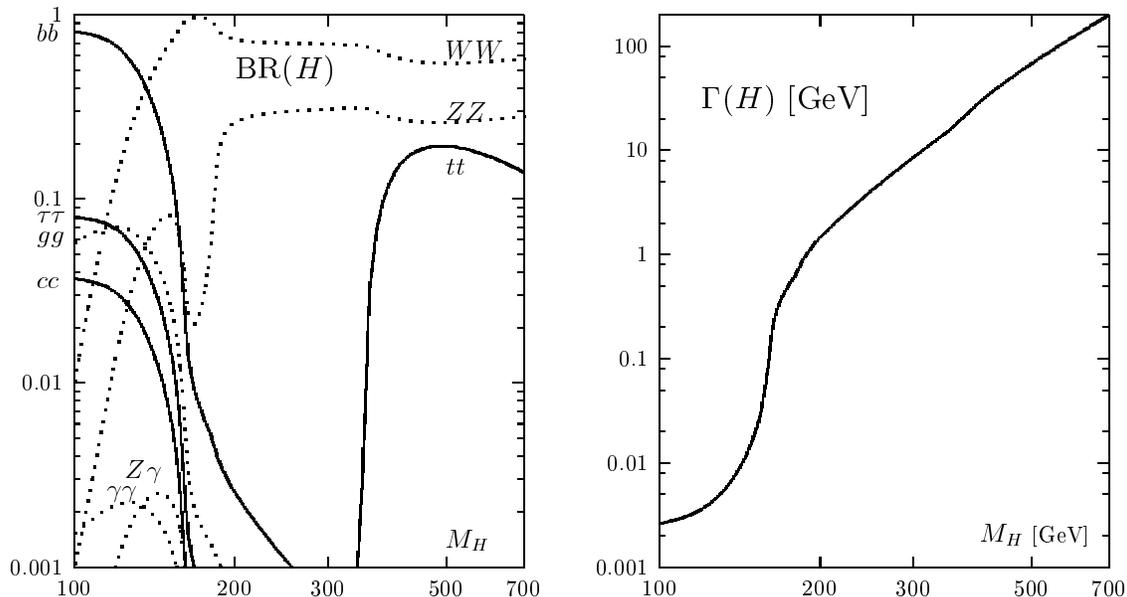,width=15cm,height=8cm}
\end{center}
\vspace*{-3mm}
\caption[]{The decay branching ratios (left) and the total decay width (right) 
of the SM Higgs boson as a function of its mass.}
\label{fig:hbrth}
\vspace*{-.1cm}
\end{figure}

\begin{figure}[htbp!]
\begin{center}
\vspace*{-2.9cm}
\hspace*{-2cm}
\epsfig{file=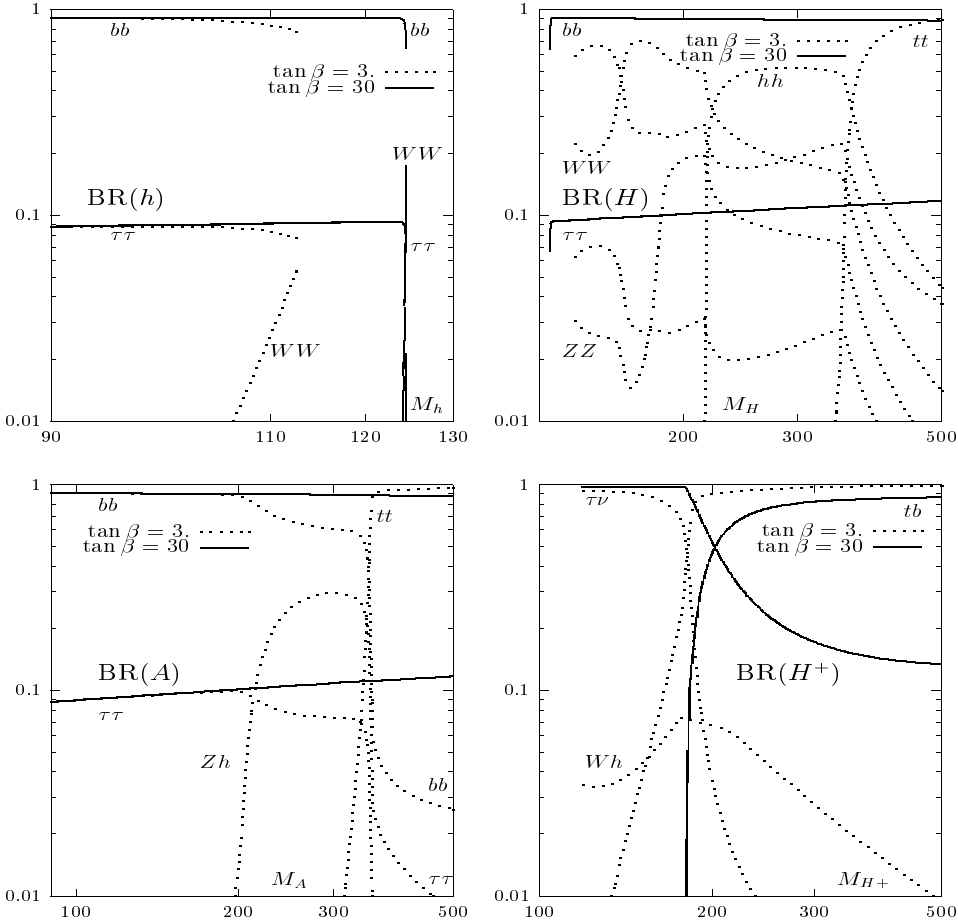,width=16.cm,height=18cm} 
\end{center}
\vspace*{-4.1cm}
\caption{Dominant MSSM Higgs bosons decay branching ratios as functions of 
the Higgs boson masses for $\tb=3$ and 30.}
\label{fig:mssmbr}
\label{fig:MSSMdecays} 
\vspace*{-.1cm}
\end{figure}

The decay pattern of the Higgs bosons of the MSSM \cite{R13} is more
complicated than in the SM and depends strongly on the value of $\tb$; see
Fig.~\ref{fig:MSSMdecays}.  

The lightest $h$ boson will decay  mainly into fermion pairs since its mass is
smaller than $\sim$~130~GeV.  This is, in general, also the dominant decay mode
of the pseudoscalar boson $A$.  For values of $\tb$ much larger than unity, the
main decay modes of the three neutral Higgs bosons are decays into $b \bar{b}$
and $\tau^+ \tau^-$ pairs with the branching ratios being of order $ \sim 90\%$
and $10\%$, respectively. For large masses, the top decay channels $H, A
\rightarrow t\bar{t}$ open up, yet for large $\tb$ these modes remain
suppressed.  If the masses are high enough, the heavy $H$ boson can decay into
gauge bosons or light $h$ boson pairs and the pseudoscalar $A$ particle into
$hZ$ final states. However, these decays are strongly suppressed for $\tb \gsim
3$--$5$ as is is suggested by the LEP2 constraints.  

The charged Higgs particles decay into fermions pairs: mainly $t\bar{b}$ and
$\tau \nu_{\tau}$ final states for $H^\pm$ masses, respectively, above and below
the $tb$ threshold.  If allowed kinematically and for small values of $\tb$, 
the $H^\pm$ bosons decay also into $hW$ final states for $\tb \lsim 5$.

Adding up the various decay modes, the widths of all five Higgs bosons remain 
very narrow. The total width of one the CP--even Higgs particle will be close 
to the SM Higgs width, while the total widths of the other Higgs particles 
will be proportional to $\tb$ and will be of the order of 10~GeV even for 
large masses.  

Other possible decay channels for the MSSM bosons, in particular the heavy $H,
A$ and $H^\pm$ states, are decays into supersymmetric particles \cite{R14}. In
addition to light sfermions, decays into charginos and neutralinos could
eventually be important if not dominant.  Decays of the lightest $h$ boson into
the lightest neutralinos (LSP) or sneutrinos can be also important, exceeding
50\% in some parts of the SUSY parameter space. These decays can render the
search for Higgs particle rather difficult, in particular at hadron colliders. 

In more general SUSY scenarii, the decays of the Higgs bosons can be much more 
complicated than in the MSSM. In particular decays of the heavy Higgses into 
gauge bosons and cascade decays into lighter Higgs bosons are still allowed.
This might render the search strategies of these particles complicated at the 
LHC. At $\ee$ colliders however, this does not lead to any difficulty to 
detect some of the particles as will be discussed later. 

\subsection*{3. Higgs Production at Hadron Colliders}

The main production mechanisms of neutral Higgs bosons in the SM at hadron 
colliders are the following processes \cite{P1}
\begin{eqnarray}
\begin{array}{lccl}
(a) & \ \ {\rm gluon-gluon~fusion} & \ \ gg  \ \ \ra & H \nonumber \\
(b) & \ \ WW/ZZ~{\rm fusion}       & \ \ VV \  \ra &  H \nonumber \\
(c) & \ \ {\rm association~with}~W/Z & \ \ q\bar{q} \ \ \ra & V + H \nonumber
\\
(d) & \ \ {\rm association~with~}Q\bar{Q} & gg,q\bar{q}\ra & Q\bar{Q}+H
\nonumber
\end{array}
\end{eqnarray}
The cross sections are shown in Fig.~6 for the LHC with $\sqrt{s}=14$ TeV and 
for the Tevatron with $\sqrt{s}=2$ TeV as functions of the Higgs boson masses; 
from Ref.~\cite{P2}.\vspace*{-.5cm}

\begin{figure}[htbp]
\hspace*{-1.5cm}
%\vspace*{-1.5cm}
\epsfig{file=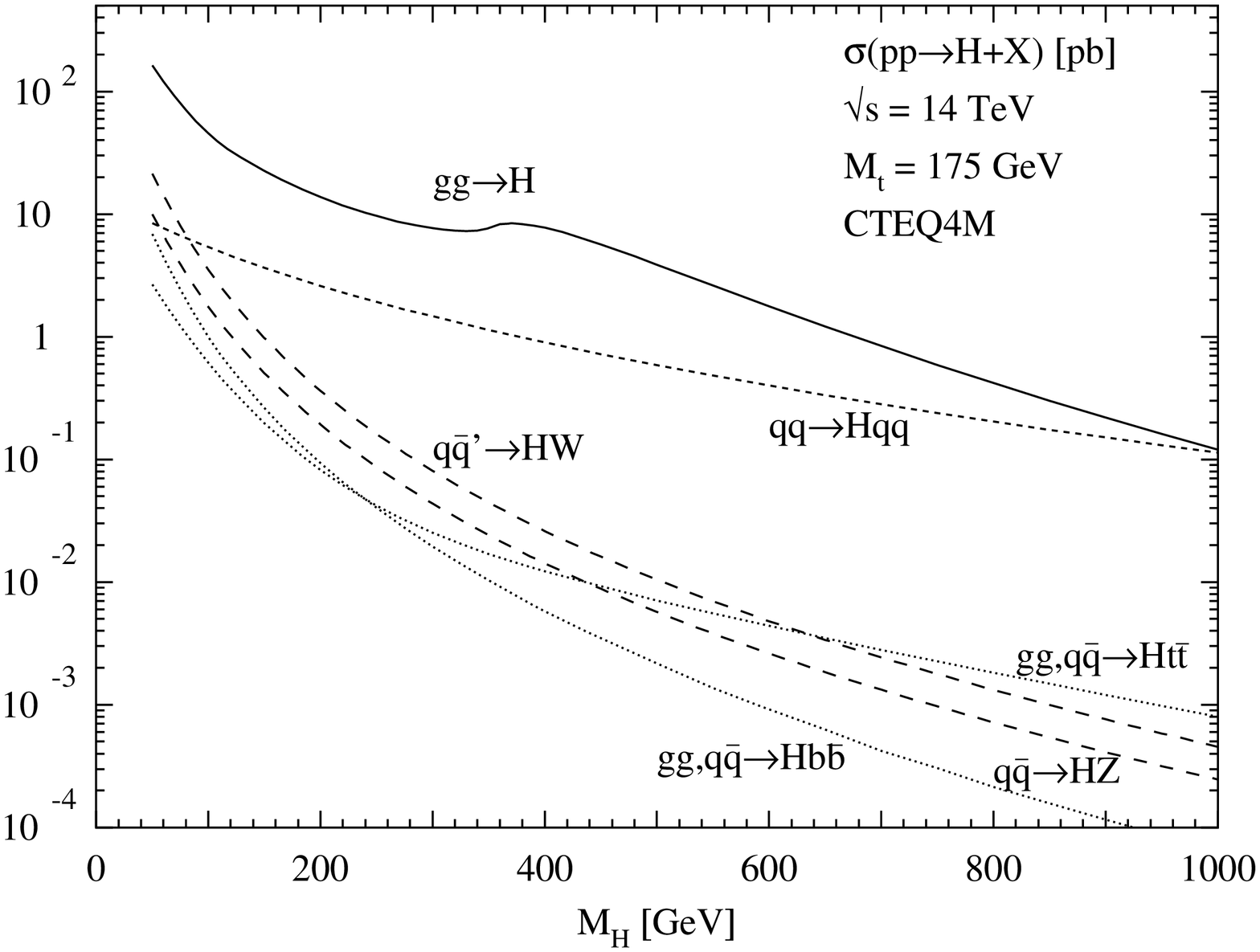,bbllx=3,bblly=3,bburx=410,bbury=420,width=5.cm,angle=-90,clip=}\hspace*{2.5cm}
\epsfig{file=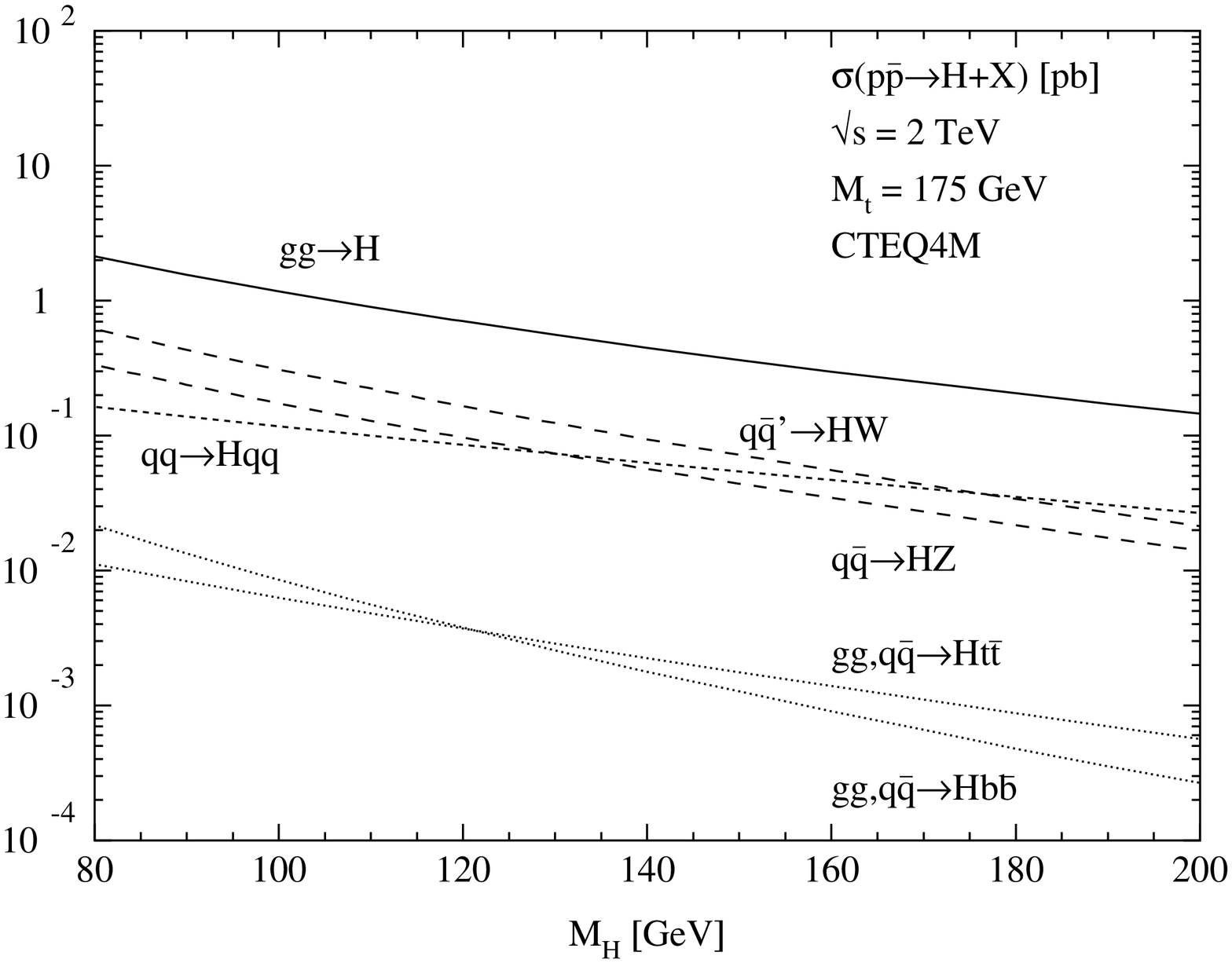,bbllx=3,bblly=3,bburx=410,bbury=420,width=5.cm,angle=-90,clip=} 
\vspace*{1.3cm}\\[.1cm]
\caption{Higgs boson production cross sections at the LHC 
(left) and the Tevatron (right) for the various mechanisms as functions of the 
Higgs mass.}
\end{figure}

At the LHC, in the interesting mass range 100 GeV $\lsim M_H \lsim 250$ GeV,
the dominant production process of the SM Higgs boson is by far the
gluon--gluon fusion mechanism [in fact it is the case of the entire Higgs mass
range] the cross section being of the order a few tens of pb. It is followed by
the $WW/ZZ$ fusion processes with a cross section of a few pb [which reaches
the level of $gg$ fusion for very large $M_H$]. The cross sections of the
associated production with $W/Z$ bosons or $t\bar{t}, b\bar{b}$ pairs are one
to two orders of magnitude smaller than the $gg$ cross section. Note that for
an integrated luminosity $\int {\cal L}=(10)~100$ fb$^{-1}$ in the low (high)
luminosity option, $\sigma=$~1 pb would correspond to $10^{4}(10^{5})$ events.

At the Tevatron, the most relevant production mechanism is the associated
production with $W/Z$ bosons, where the cross section is slightly less than a
picobarn for $M_H \sim 120$ GeV, leading to  $\sim 10^4$ Higgs events for a
luminosity $\int {\cal L}=20$ fb$^{-1}$. The $WW/ZZ$ fusion cross sections are
slightly smaller for $M_H \lsim 150$ GeV, while the cross sections for
associated production with $t\bar{t}$ or $b\bar{b}$ pairs are rather low.  The
$gg$ fusion mechanism has the largest cross section but suffers from a huge QCD
two--jet background.  

The next--to--leading order QCD corrections should be taken into account in the
$gg$ fusion processes where they are large, leading to an increase of the
production cross sections by a factor of up to two \cite{P3}. For the other
processes, the QCD radiative corrections are relatively smaller \cite{P4}: for
the associated production with gauge bosons, the corrections [which can be
inferred from the Drell--Yan $W/Z$ production] are at the level of $10\%$,
while in the case of the vector boson fusion processes, they are at the level
of $30\%$. For the associated production with top quarks, the NLO corrections
alter the cross section by $\sim 20\%$ if the scale is chosen properly. In all
these production processes, the theoretical uncertainty, from the remaining 
scale dependence and from the choice of different sets of parton densities, can
be estimated as being of the order of $\sim 20$--$30\%$.  

The signals which are best suited to identify the produced Higgs particles at
the Tevatron and at the LHC have been studied in great detail in
Refs.~\cite{R8,R9}, respectively. I briefly summarize below the main
conclusions of these studies.

At the Tevatron Run II, the associated production with $W/Z$ bosons with the
latter decaying leptonically lead to several distinct signatures in which a
signal can be observed with sufficient integrated luminosity. In the low Higgs
mass range, $M_H \lsim 130$ GeV, the Higgs will mainly decay into $b\bar{b}$
pairs and the most sensitive signatures are $\ell \nu b\bar{b}$, $\nu\bar{\nu}
b\bar{b}$, and $\ell^+\ell^- b\bar{b}$. Hadronic decays of the $W$ and $Z$ lead
to the $q\bar{q}b\bar{b}$ final state and cannot be used since they  suffers
from large backgrounds from QCD multi-jet production. In the high Higgs mass
range, $M_H \gsim 130$ GeV, the dominant decay is $H \to WW^*$ and the
signature $\ell^\pm \ell^\pm jj$ from three vector boson final states can be
used. In addition, one can use the final state $\ell^+ \ell^- \nu\bar{\nu}$
with the Higgs boson produced in $gg$ fusion.  

The required luminosity to discover or exclude a SM Higgs boson, combining all
channels in both D0 and CDF experiments, is shown in Fig.~7 as a function of
$M_H$ \cite{R8}.  With 15 fb$^{-1}$ luminosity, a 5$\sigma$ signal can be
achieved for $M_H \lsim 120$ GeV, while a Higgs boson with a mass $M_H \lsim
190$ GeV can be excluded at the 95\% confidence level.  
 
\begin{figure}[htbp]
\vspace*{-.5cm}
\begin{center}
\includegraphics[width=10.cm]{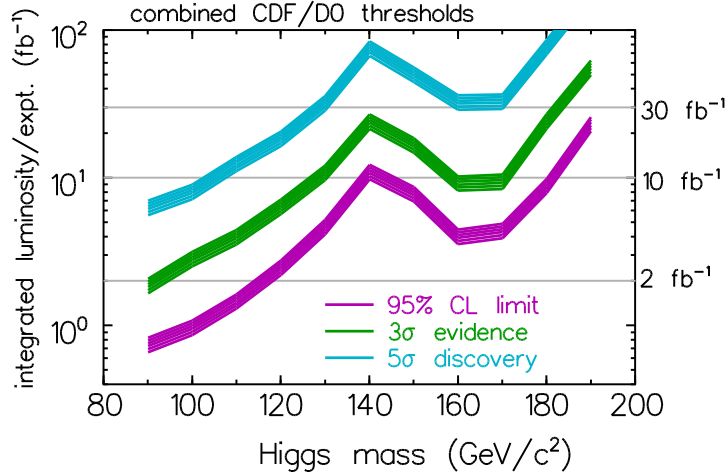}
\vspace*{-0.1cm}
  \caption{The integrated luminosity required per experiment, to
            either exclude a SM Higgs boson at 95\% CL or discover it at the
            $3\sigma$ or $5\sigma$ level, as a function of $M_H$.} 
\end{center}
\vspace*{-0.3cm}
\end{figure}

Let us now turn to the signatures which can be used at the LHC. A discovery
with a significance larger than 5$\sigma$ can be obtained used various 
channels; see Figure 8. 

In the high mass range, $M_H \gsim 130$ GeV, the signal consists of the
so--called ``gold--plated" events $H \ra Z Z^{(*)} \ra 4 \ell^\pm$ with
$\ell=e,\mu$. The backgrounds, mostly $pp \ra ZZ^{(*)}, Z \gamma^*$ for the
irreducible background and $t \bar{t} \ra WWb \bar{b}$ and $Zb \bar{b}$ for the
reducible one, are relatively small. One can probe Higgs boson masses up to
${\cal O}$(500~GeV) with a luminosity $\int {\cal L}= 100 $~fb$^{-1}$.  The
channels $H \to ZZ \to \nu\bar{\nu} \ell^+ \ell^-$ and $H \to WW \to \nu \ell
jj$, which have larger rates, allow to extend the reach to $M_H \sim 1$ TeV. The
$H \ra WW^{(*)} \to \nu \bar{\nu} \ell^+ \ell^-$ [with $H$ produced in
$gg$ fusion, and to a lesser extent, in association with $W$ bosons] decay
channel is very useful in the range 130 GeV $\lsim M_H \lsim 180$ GeV,
where BR($H \to ZZ^*$) is too small, despite of the large background from $WW$
and $t\bar{t}$ production.

For the ``low mass" range, the situation is more complicated. The branching
ratios for $H\ra ZZ^*, WW^*$ are too small and due to the huge QCD jet
background, the dominant mode $H\ra b\bar{b}$ is practically useless. One has
then to rely on the rare $\gamma \gamma$ decay mode with a branching ratio of
${\cal O}(10^{-3})$, where the Higgs boson is produced in the $gg$ fusion 
and the associated $WH$ and $Ht\bar{t}$ processes. A $5\sigma$ discovery can 
be obtained with a luminosity $\int {\cal L}= 100$~fb$^{-1}$, despite of the 
formidable backgrounds. Finally, in the very low mass range, $M_H \sim 115$ 
GeV, the channel $pp \to t\bar{t}H$ with $H \to b\bar{b}$ can be used.  

\begin{figure}[hbt]
\vspace*{-3mm}
\begin{center}
\epsfig{file=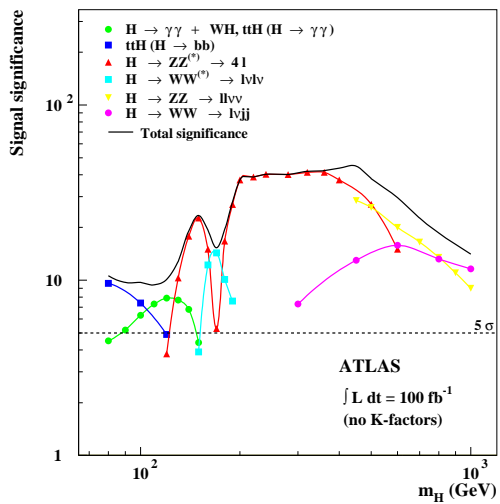,width=6.5cm}\hspace*{4mm}
\end{center}
\caption{Significance for the SM Higgs boson discovery in various channels at 
the LHC with a high luminosity as a function of the Higgs mass.}
\vspace*{-.2cm}
\end{figure}

In the MSSM, the production processes for the neutral CP--even Higgs particles
are practically the same as for the SM Higgs. However, for large $\tb$ values,
one has to take the $b$ quark [whose couplings are strongly enhanced] into
account: its loop contributions in the $gg$ fusion process [and also the extra
contributions from squarks loops, which however decouple for high squark
masses] and associated production with $b\bar{b}$ pairs. The cross sections for
the associated production with $t\bar{t}$ pairs and $W/Z$ bosons and the
$WW/ZZ$ fusion processes, are suppressed for at least one of the particles
because of the coupling suppression.  Because of CP--invariance, the
pseudoscalar $A$ boson can be produced only in the $gg$ fusion and in
association with heavy quarks [associated production with a CP--even Higgs
particle, $pp \to A+h/H$, is also possible but the cross section is too small].
For high enough $\tb$ values and for $M_A \gsim (\lsim) 130$ GeV, the
$gg/q\bar{q}\ra b\bar{b}+A/H(h)$ and $gg \to A/H(h)$ processes become the
dominant production mechanisms.  

The charged Higgs particles, if lighter than the top quark, can be
accessible in the decays $t \ra H^+b$ with $H^-\ra \tau \nu_\tau$, leading to
a surplus of $\tau$ events mimicking a breaking of $\tau$ versus $e,\mu$ 
universality. The $H^\pm$ particles can also be produced directly in the
[properly combined] processes $gb \to tH^-$ or $qq/ gg \to H^-t \bar{b}$. 
 
\begin{figure}[htbp]
\vspace*{-.7cm}
\hspace*{-1.5cm}
\epsfig{file=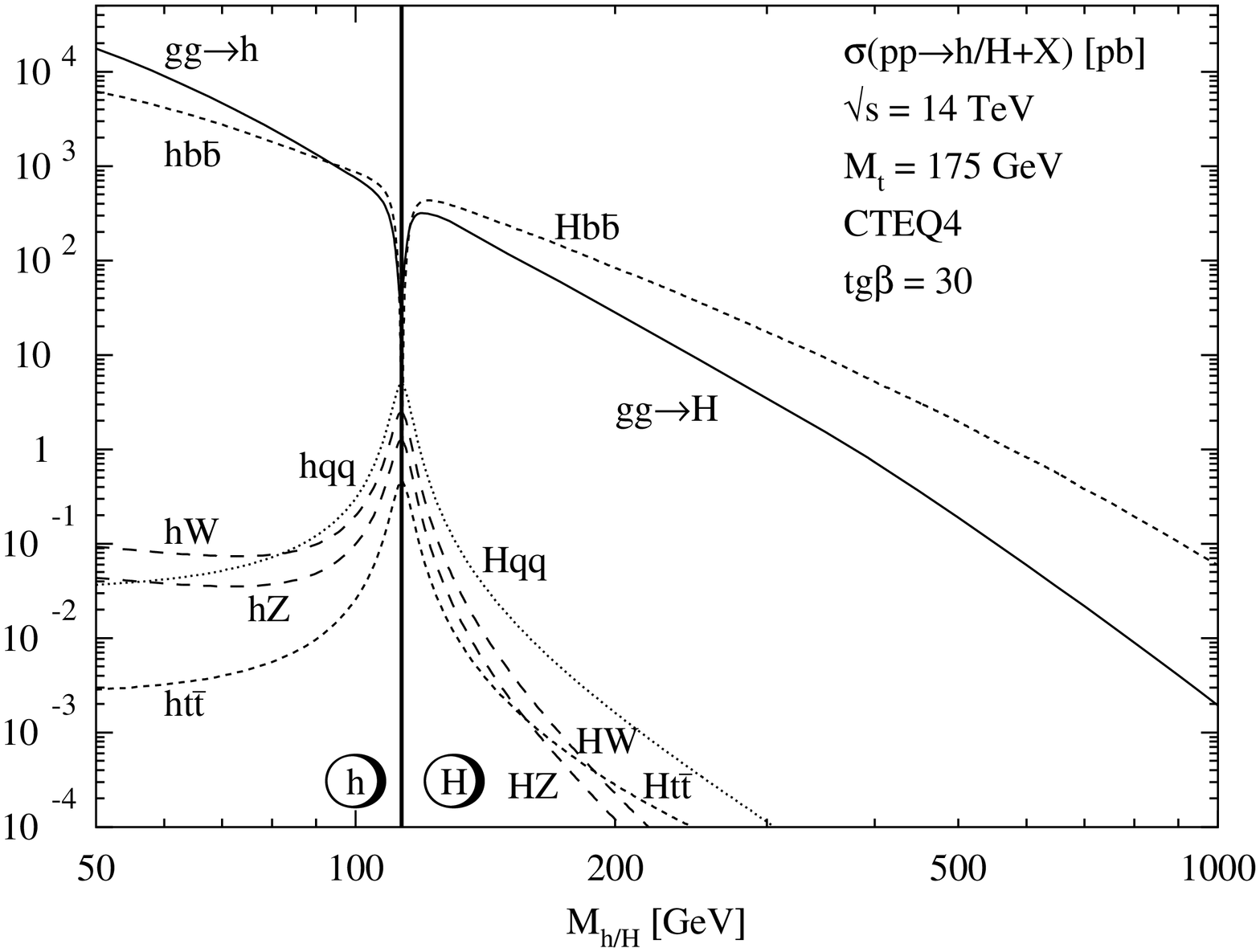,bbllx=3,bblly=3,bburx=410,bbury=420,width=5.cm,angle=-90,clip=}\hspace*{2.8cm}
\epsfig{file=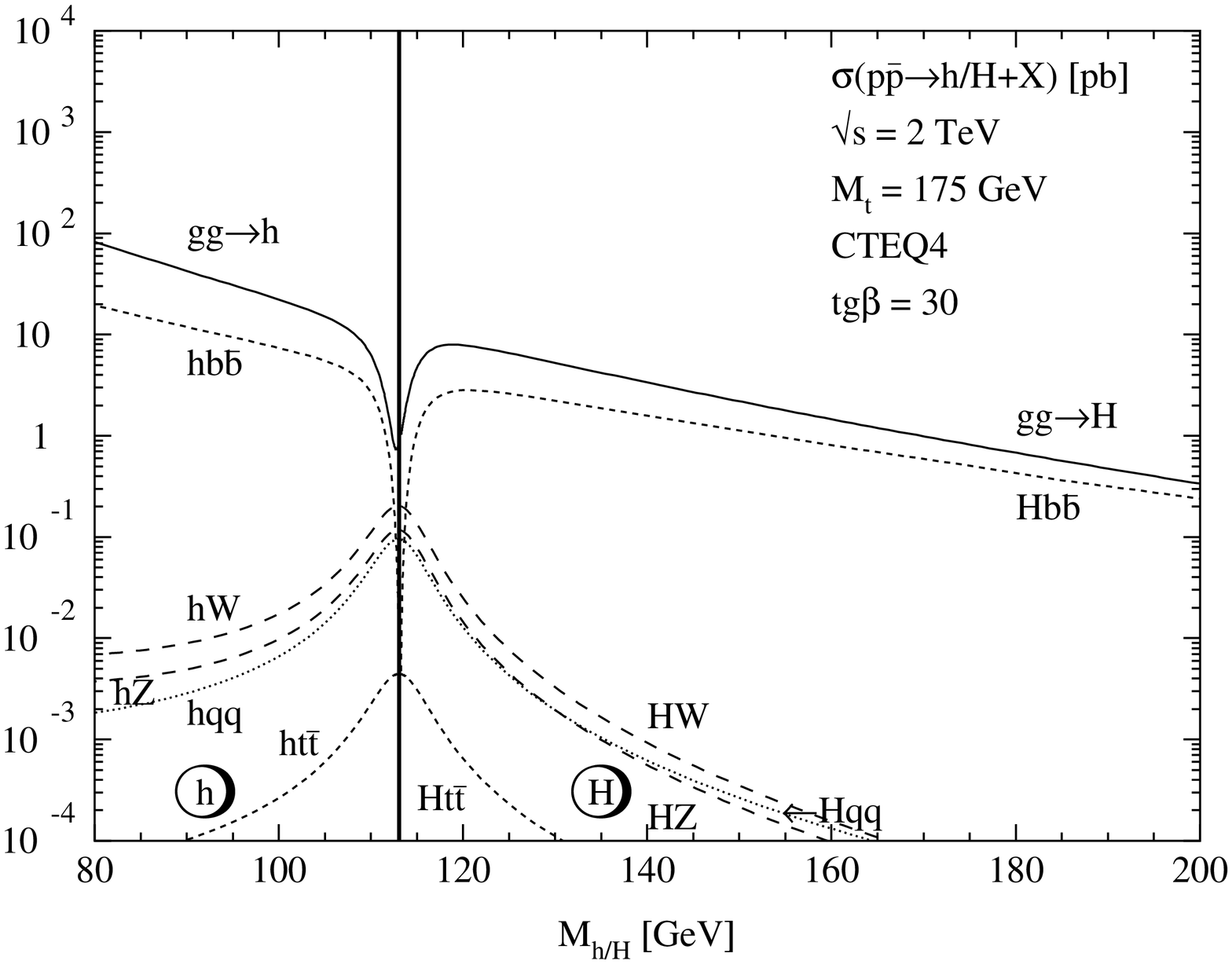,bbllx=3,bblly=3,bburx=410,bbury=420,width=5.cm,angle=-90,clip=}\\[1.4cm] 
%\vspace*{1.3cm}
\caption{CP--even Higgs production cross sections at  LHC 
(left) and Tevatron (right) for the various mechanisms as a function of the 
Higgs masses for $\tb=30$.}
\vspace*{-2mm}
\end{figure}

The cross sections for the production of the CP--even Higgs particles 
are shown in Fig.~9 for the Tevatron and LHC for $\tb=30$ [the cross sections
for $A$ production are roughly equal to the one of the $h(H)$ particle 
in the low (high) mass range]; from Ref.~\cite{P2}. The various detection 
signals can be briefly summarized as follows [see also Fig.~10]:

$(i)$ Since the lightest Higgs boson mass is always smaller than $\sim 130 $
GeV, the $WW$ and $ZZ$ signals cannot be used. Furthermore, the $hWW
(h\bar{b}b)$ coupling is suppressed (enhanced) leading to a smaller $\gamma
\gamma$ branching ratio than in the SM, making the search in this channel more
difficult. If $M_h$ is close to its maximum value, $h$ has SM like couplings
and the situation is similar to the SM case with $M_H \sim $ 100--130 GeV.  
  
$(ii)$ Since $A$ has no tree--level couplings to gauge bosons and since the
couplings of $H$ are strongly suppressed, the gold--plated $ZZ$ signal is lost
[for $H$ it survives only for small $\tb$ values, provided that $M_H<2m_t$]. In
addition, the $A/ H \ra \gamma \gamma$ signals cannot be used since the
branching ratios are suppressed.  One has then to rely on the $A/H  \ra \tau^+
\tau^-$ or even $\mu^+ \mu^-$ channels for large $\tb$ values.  [The decays $H
\ra hh \ra b \bar{b}b\bar{b}$, $A \ra hZ \ra Zb \bar{b}$ and $H/A \ra t\bar{t}$
have too small rates in view of the LEP2 constraints]. 

$(iii)$ Light $H^\pm$ particles can be observed in the decays $t \ra H^+b$ with
$H^-\ra \tau \nu_\tau$ where masses up to $\sim 150$ GeV can be probed. The 
mass reach can be extended up to a few hundred GeV for $\tb \gg 1$, by 
considering the processes \cite{P5} $gb \to t H^-$ and $gg \ra t \bar{b} H^-$ 
with the decays $H^-\ra \tau \nu_\tau$ [using $\tau$ polarization] or 
$\bar{t}b$. 

$(iv)$ All the previous discussion assumes that Higgs decays into SUSY 
particles are kinematically inaccessible. This seems to be  unlikely since at 
least the decays of the heavier $H,A$ and $H^\pm$ particles into charginos 
and neutralinos should be possible \cite{R14}. Preliminary analyses show that 
decays into neutralino/chargino final states $H/A \to \chi_2^0 \chi_2^0 \to 4 
\ell^\pm X$ and $H^\pm \to \chi_2^0 \chi_1^\pm \to 3 \ell^\pm X $ can be 
detected in some cases \cite{P6}. It could also be possible that 
the lighter $h$ decays invisibly into the lightest neutralinos or sneutrinos.  
If this scenario is realized, the discovery of these Higgs particles will be
more challenging. Preliminary analyses for the 2001 les Houches Workshop 
in Ref.~\cite{R9} show however, that an invisibly decaying Higgs boson could be 
detected in the $WW$ fusion process.  

$(v)$ If top squarks are light enough, their contribution to the $gg$ fusion
mechanism and to the $\gamma \gamma$ decay [here, this is also the case for
light charginos] should be taken into account. In the large mixing scenario,
stops can be rather light and couple strongly to the $h$ boson, leading to a
possibly strong suppression of the product $\sigma( gg \to h)\times {\rm BR} (h
\to \gamma \gamma)$ \cite{P7}. However, in this case, the associated production
of the $h$ boson with top squarks is possible and the cross sections would be
rather sizeable \cite{P8}.  

$(vi)$ MSSM Higgs boson detection from the cascade decays of Supersymmetric
particles, originating from squark and gluino production, are also possible. 
In particular, the production of the lighter $h$ boson from the decays of the
next-to-lightest neutralino and the production of $H^\pm$ from the decays of
the heavier chargino/neutralino states into the lighter ones have been
discussed; see Ref.~\cite{P9} for instance.

\begin{figure}[hbtp]
\vspace*{-1cm}
\begin{center}
\epsfig{file=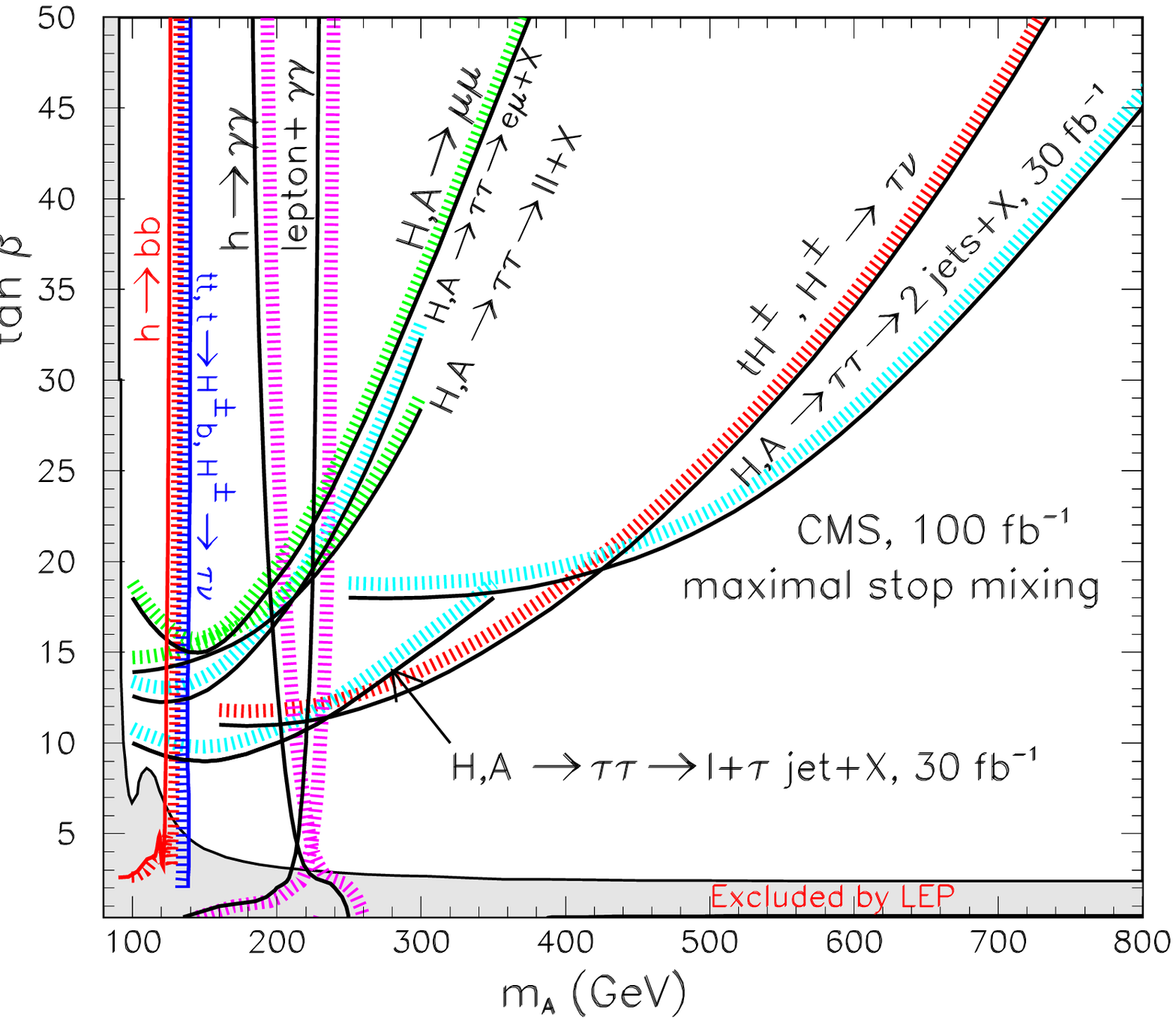,width=7.5cm,height=7.8cm}\hspace*{-3mm}
\epsfig{file=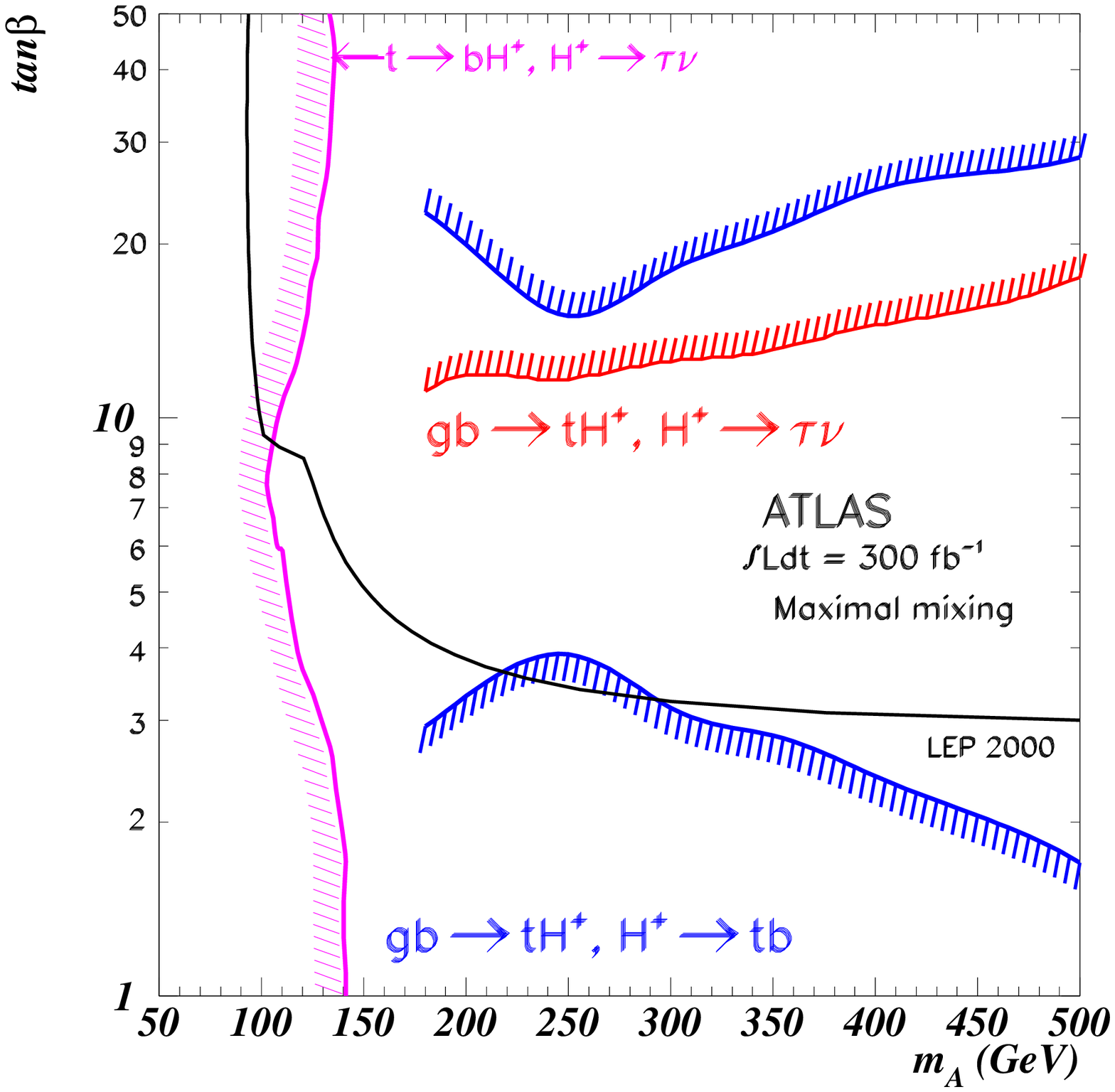,width=7.8cm,height=7.8cm}
\end{center}
\vspace*{-5mm}
\caption{Neutral (left) and charged (right) MSSM Higgs boson discovery at 
the LHC in various channels in the $(M_A, \tb)$ plane with a high luminosity.}
\end{figure}

At the Tevatron Run II, the search for the CP--even $h$ and $H$ bosons will be
more difficult than in the SM because of the reduced couplings to gauge bosons,
unless one of the Higgs particles is SM--like. However, associated production
with $b\bar{b}$ pairs, $pp \to b\bar{b}+A/h(H)$ in the low (high) $M_A$ range
with the Higgs bosons decaying into $b\bar{b}$ pairs, might lead to a visible
signal for rather large $\tb$ values and $M_A$ values below the 200 GeV range. 
The $H^\pm$ boson would be also accessible in top quark decays for large or
small values of $\tb$, for which the branching ratio BR$(t \to H^+b)$ is large
enough,

In more general SUSY extensions of the SM, the Higgs spectrum can be much more
complicated than in the MSSM. While the production mechanisms will probably
remain the same, the production cross sections can be different. The decays
signatures could also be much more complicated than in the MSSM. This would
make the search for the Higgs bosons of these extensions very challenging. 
There are no detailed simulations which have been performed on this issue yet. 
A preliminary theoretical analysis \cite{P10} in the context of the NMSSM, i.e.
the MSSM supplemented by one Higgs singlet, has been devoted to the
observability of at least one Higgs boson at the LHC with 300 fb$^{-1}$
integrated luminosity, taking the present LEP2 constraints into account and
making use of the $WW$ fusion mechanism. It concludes that the LHC will
discover at least one NMSSM Higgs boson unless there are large branching ratios
for decays to SUSY particles and/or to other Higgs bosons. This analysis needs
to be confirmed by detailed (experimental) simulations.  
 
\subsection*{4. Higgs Production at $e^+ e^-$ Colliders}

At $\ee$ linear colliders operating in the 300--800 GeV energy range,  the main
production mechanisms for SM--like Higgs particles are \cite{E1}
\begin{eqnarray} 
\begin{array}{lccl} 
(a)  & \ \ {\rm bremsstrahlung \ process} & \ \ \ee & \ra (Z) \ra Z+H \non \\ 
(b)  & \ \ WW \ {\rm fusion \ process} & \ \ \ee & \ra \bar{\nu} \ \nu \ (WW) 
\ra \bar{\nu} \ \nu \ + H \non \\ 
(c)  & \ \ ZZ \ {\rm fusion \ process} & \ \ \ee & \ra e^+ e^- (ZZ) \ra 
e^+ e^- + H \non \\ 
(d)  & \ \ {\rm radiation~off~tops} & \ \ \ee & \ra (\gamma,Z) \ra t\bar{t}+H 
\non \end{array} 
\end{eqnarray} 
The Higgs--strahlung cross section scales as $1/s$ and therefore dominates at
low energies while the $WW$ fusion mechanism  has a cross section which rises
like $\log(s/M_H^2)$ and dominates at high energies.  At $\sqrt{s} \sim 500$
GeV, the two processes have approximately the same cross sections, ${\cal O}
(100~{\rm fb})$ for the interesting range 100 GeV $\lsim M_H \lsim$ 200 GeV, as
shown in Fig.~11.  With an integrated luminosity $\int {\cal L} \sim 500$
fb$^{-1}$, as expected for instance at the TESLA machine \cite{R12},
approximately 25.000 events per year can be collected in each channel for a
Higgs boson with a mass $M_H \sim 150$ GeV. This sample is more than enough to
discover the Higgs boson and to study its properties in detail.  

The $ZZ$ fusion mechanism has a cross section which is one order of magnitude
smaller than $WW$ fusion, a result of the smaller neutral couplings compared to
charged current couplings. The associated production with top quarks has a very
small cross section at $\sqrt{s}=500$ GeV due to the phase space suppression
but at $\sqrt{s}=1$ TeV it can reach the level of a few femtobarn.  Despite of
the small production cross sections, shown in Fig.~11 as a function of
$\sqrt{s}$ for $M_H=120$ GeV, these processes will be very useful when it comes
to study the Higgs properties as will be discussed later.  The cross section
for the double Higgs production in the strahlung process \cite{E2}, $\ee \to
HHZ$, also shown in Fig.~11 is at the level of a fraction of a femtobarn and
can be used to extract the Higgs self--coupling.

\begin{figure}[hbtp]
\begin{center}
\vspace*{-5.1cm}
\hspace*{-1.2cm}
\epsfig{file=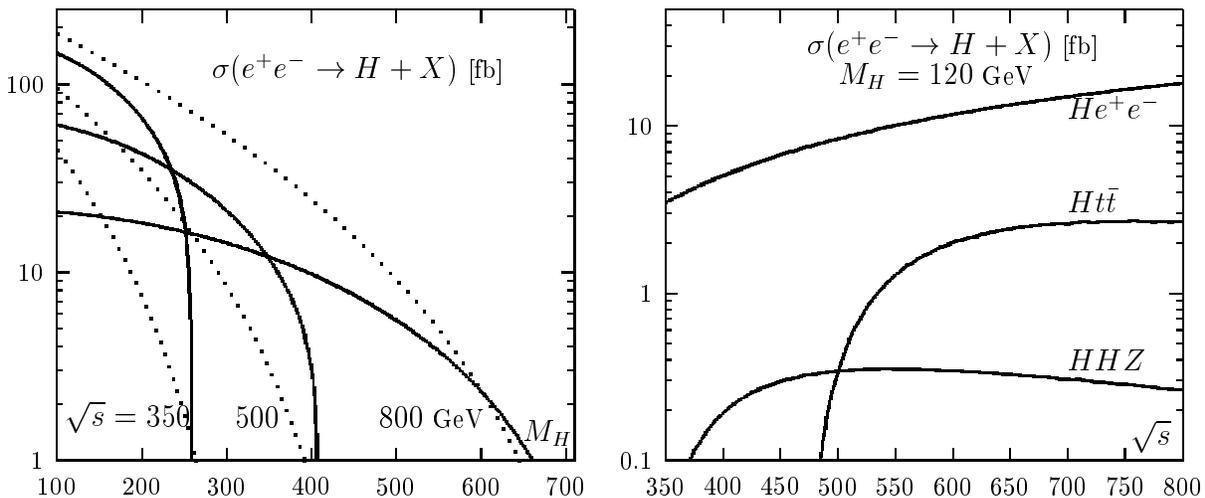,width=20.5cm}
\vspace*{-18.9cm}
\end{center}
\caption[]{Production cross sections of the SM Higgs boson in $e^+ e^-$ in the 
main processes with $\sqrt{s}=350,500$ and 800 GeV as a function of $M_H$ (left)
and in higher order process as a function of $\sqrt{s}$ for $M_H=120$ GeV
(right).}
\vspace*{-0.3cm}
\end{figure}

In the Higgs--strahlung process, the recoiling $Z$ boson [which can be tagged
through its clean $\mu^+ \mu^-$ decay] is mono--energetic and $M_H$ 
can be derived from the energy of the $Z$ if the initial $e^+$ and $e^-$ beam
energies are sharp [beamstrahlung, which smears out the c.m. energy should thus
be suppressed as strongly as possible, and this is already the case for machine
designs such as TESLA]. Therefore, it will be easy to separate the signal from
the backgrounds. For low Higgs masses, $M_H \lsim 130$ GeV, the main background
will be $\ee \ra ZZ$. The cross section is large, but it can be reduced by
cutting out the forward and backward directions [the process is mediated by
$t$--channel $e$ exchange] and by selecting $b\bar{b}$ final states by means of
$\mu$--vertex detectors [while the Higgs decays almost exclusively into
$b\bar{b}$ in this mass range, BR$(Z \ra b\bar{b}$) is small, $\sim 15\%$]. The
background from single $Z$ production, $\ee \ra Zq\bar{q}$, is small and can be
further reduced by flavor tagging. In the mass range where the decay $H \ra
WW^*$ is dominant, the main background is triple gauge boson production and is
suppressed by two powers of the electroweak coupling. 

The $WW$ fusion mechanism offers a complementary production channel. For small
$M_H$, the main backgrounds are single $W$ production, $\ee \ra e^\pm W^\mp
\nu$ $[W \ra q\bar{q}$ and the $e^\pm$ escape detection] and $WW$ fusion into a
$Z$ boson, $\ee \ra \nu \bar{\nu}Z$, which have cross sections 60 and 3 times
larger than the signal, respectively. Cuts on the rapidity spread, the energy
and momentum distribution of the two jets in the final state [as well as flavor
tagging for small $M_H$] will suppress these background events. 

It has been shown in detailed simulations \cite{R12} that only a few fb$^{-1}$
of integrated luminosity are needed to obtain a 5$\sigma$ signal for a Higgs
boson with a mass $M_H \lsim 140$ GeV at a 500 GeV collider [in fact, in this
case, it is better to go to lower energies where the cross section is larger],
even if it decays invisibly [as it could happen in SUSY models for instance].
Higgs bosons with masses up to $M_H \sim 400$ GeV can be discovered at the
5$\sigma$ level, in both the strahlung and fusion processes at an energy of 500
GeV and with a luminosity of 500 fb$^{-1}$. For even higher masses, one needs
to increase the c.m. energy of the collider, and as a rule of thumb, Higgs
masses up to $\sim 80\% \sqrt{s}$ can be probed. This means than a $\sim 1$ 
TeV collider will be needed to probe the entire SM Higgs mass range.  

An even stronger case for $\ee$ colliders in the 300--800 GeV energy range is
made by the MSSM. In $\ee$ collisions \cite{E3}, besides the usual
bremsstrahlung and fusion processes for $h$ and $H$ production, the neutral
Higgs particles can also be produced pairwise: $\ee \ra A + h/H$.  The cross
sections for the bremsstrahlung and the pair production as well as the cross
sections for the production of $h$ and $H$ are mutually complementary, coming
either with a coefficient $\sin^2(\beta- \alpha)$ or $\cos^2(\beta -\alpha)$;
see Fig.~12. The cross section for $hZ$ production is large for large values of
$M_h$, being of ${\cal O}(100$ fb) at $\sqrt{s}=350$ GeV; by contrast, the
cross section for $HZ$ is large for light $h$ [implying small $M_H$].  In major
parts of the parameter space, the signals consist of a $Z$ boson and $b\bar{b}$
or $\tau^+ \tau^-$ pairs, which is easy to separate from the main background,
$\ee \ra ZZ$ [in particular with $b$--tagging]. For the associated production,
the situation is opposite: the cross section for $Ah$ is large for light $h$
whereas $AH$ production is preferred in the complementary region.  The signals
consists mostly of four $b$ quarks in the final state, requiring efficient
$b\bar{b}$ quark tagging; mass constraints help to eliminate the QCD jets and
$ZZ$ backgrounds. The CP--even Higgs particles can also be searched for in the
$WW$ and $ZZ$ fusion mechanisms.

\begin{figure}[hbtp]
\begin{center}
\vspace*{-5.1cm}
\hspace*{-1.2cm}
\epsfig{file=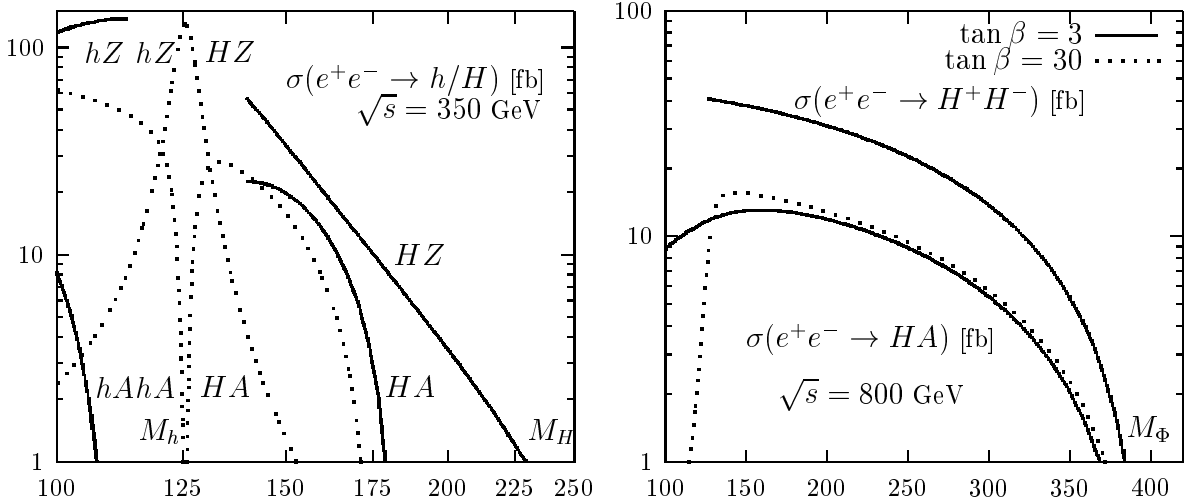,width=20.5cm}
\vspace*{-18.9cm}
\end{center}
\caption[]{Production cross sections of the MSSM Higgs bosons in $e^+ e^-$
as functions of the masses: $h,H$ production at $\sqrt{s}=350$ GeV (left) 
and $HA,H^+H^-$ production at  $\sqrt{s}=800$ GeV (right); the dotted (full)
lines are for $\tb=30(3)$.} 
\vspace*{-0.3cm}
\end{figure}

In $\ee$ collisions, charged Higgs bosons can be produced pairwise, $\ee \ra
H^+H^-$, through $\gamma,Z$ exchange. The cross section depends only on the
charged Higgs mass; it is large almost up to $M_{H^\pm} \sim \sqrt{s}/2$. 
$H^\pm$ bosons can also be produced in top decays as at hadron colliders; in
the range $ 1 < \tb < m_t/m_b$, the $t \ra H^+b$ branching ratio and the
$t\bar{t}$ production cross sections are large enough to allow for their
detection in this mode.  

The discussion on the MSSM Higgs production at $\ee$ linear colliders [not
mentioning yet the $\gamma \gamma$ option of the collider] can be summarized in 
the following points \cite{R12}: 

$i)$ The Higgs boson $h$ can be detected in the entire range of the MSSM
parameter space, either through the bremsstrahlung process or pair
production; in fact, this conclusion holds true even at a c.m. energy of 300
GeV and with a luminosity of a few fb$^{-1}$.  

$ii)$ All SUSY Higgs bosons can be discovered at an $\ee$ collider if the $H,A$
and $H^{\pm}$ masses are less than the beam energy; for higher masses, one
simply has to increase $\sqrt{s}$. 

$iii)$ Even if the decay modes of the Higgs bosons are very complicated [e.g.
they decay invisibly] missing mass techniques allow for their detection.  

$iv)$ The additional associated production processes with  $t\bar{t}$ and
$b\bar{b}$ allow for the measurement of the Yukawa couplings. In particular,
$\ee \to b\bar{b}+H/A$ for high $\tb$ values allows for the determination of 
this important parameter for low $M_A$ values \cite{E4}. 

In extensions of the MSSM, the Higgs production processes are as the ones above
but the phenomenological analyses are more involved since there is more 
freedom in the choice of parameters. However, even if the Higgs sector is 
extremely complicated, there is always a light Higgs boson which has sizeable
couplings to the $Z$ boson. This Higgs particle can be thus produced in the
strahlung process, $\ee \ra Z+$``$h$", and using the missing mass technique this
``$h$" particle can be detected. Recently a ``no--loose theorem" has been 
proposed \cite{E5}:  a Higgs boson in SUSY theories can be always detected
at a 500 GeV $\ee$ collider with a luminosity of $\int {\cal L} \sim 500$ fb
$^{-1}$ in the strahlung process, regardless of the complexity of the Higgs
sector of the theory and of the decays of the Higgs boson. \bigskip

Finally, future linear colliders can be turned to $\gamma \gamma$ colliders, in
which the photon beams are generated by Compton back--scattering of laser
light; c.m.~energies of the order of 80\% of the $\ee$ collider energy and
integrated luminosities $\int {\cal L} \sim 100$ fb, as well as a high degree
of longitudinal photon polarization can be reached at these colliders
\cite{E6}. \

Tuning the maximum of the $\gamma \gamma$ spectrum to the value of the Higgs
boson mass, the Higgs particles can be formed as $s$--channel resonances,
$\gamma \gamma \to $ Higgs, decaying mostly into $b\bar{b}$ pairs \cite{E7}.
The main background, $\gamma \gamma \to b\bar{b}$, can be suppressed by
choosing proper helicities for the initial $e^\pm$ and laser photons which
maximizes the signal cross section, and eliminating the gluon radiation by
taking into account only two--jet events.  Clear signals can be obtained which
allow the measurement of the Higgs couplings to photons, which are mediated by
loops possibly involving new particles. In addition, in the MSSM, $\gamma
\gamma$ colliders allow to extend the reach for the heavy $H,A$ bosons compared
to the $\ee$ option \cite{E8}.

\subsection*{5. Determination of the properties of a SM--like Higgs boson} 

Once the Higgs boson is found it will be of great importance to explore all its
fundamental properties. This can be done at great details in the clean
environment of $\ee$ linear colliders \cite{R12,M1}: the Higgs mass, the spin 
and parity quantum numbers and the couplings to fermions, gauge bosons and the
self--couplings can measured. Some precision measurements, in particular for
the mass and width,  can also be performed at the LHC with high--luminosity 
\cite{R9,R10,M2}. In the following we will summarize these features in the case
of the SM Higgs boson; some of this discussion can be of course extended to the
the lightest MSSM Higgs particle. We will rely on Refs.~\cite{R10,M2} 
and \cite{R12,M1} for the LHC and TESLA analyses, respectively, where the 
references for the original studies can be found. 

\subsubsection*{5.1 Studies at $\ee$ Colliders}

$\bullet$ The measurement of the recoil $\ee$ or $\mu^+ \mu^-$ mass in the
Higgs--strahlung process, $\ee \ra ZH\ra He^+e^-$ and $H\mu^+ \mu^-$, allows a
very good determination of the Higgs boson mass. At $\sqrt{s}=350$ GeV and with
a luminosity of $\int {\cal L}= 500$ fb$^{-1}$, a precision of $ \Delta M_H
\sim 70$ MeV can be reached for a Higgs boson mass of $M_H \sim 120$ GeV.  The
precision can be increased to $\Delta M_H \sim 40$ MeV by using in addition the
hadronic decays of the $Z$ boson [which have more statistics]. Accuracies of
the order of $\Delta M_H \sim 80$ MeV can also be reached for $M_H=150$ and 180
GeV when the Higgs decays mostly into gauge bosons. This one per mile accuracy
on $M_H$ can be very important, especially in the MSSM where it allow to
strongly constrain the other parameters of the model.  

$\bullet$ The angular distribution of the $Z/H$ in the Higgs--strahlung process
is sensitive to the spin--zero of the Higgs particle: at high--energies the $Z$
is longitudinally polarized and the distribution follows the $\sim
\sin^2\theta$ law which unambiguously characterizes the production of a
$J^P=0^+$ particle. The spin--parity quantum numbers of the Higgs bosons can
also be checked experimentally by looking at correlations in the production
$\ee \ra HZ \ra 4f$ or decay $H \ra WW^* \ra 4f$ processes, as well as in the
channel $H \ra \tau^+ \tau^-$ for $M_H \lsim 140$ GeV. An unambiguous test of
the CP nature of the Higgs bosons can be made in the process $\ee \ra t
\bar{t}H$ or at laser photon colliders in the loop induced process $\gamma
\gamma \ra H$.  

$\bullet$ The masses of the gauge bosons are generated through the Higgs
mechanism and the Higgs couplings to these particles are proportional to their
masses.  This fundamental prediction has to be verified experimentally.  The
Higgs couplings to $ZZ/WW$ bosons can be directly determined by measuring the
production cross sections in the bremsstrahlung and the fusion processes.  In
the $\ee \ra H \ell^+ \ell^-$ and $H\nu \bar{\nu}$ processes, the total cross 
section can be measured with a precision less than $\sim 3\%$ at $\sqrt{s}\sim
500$ GeV and with $\int {\cal L}= 500$ fb$^{-1}$. This leads to an accuracy of 
$\lsim 1.5\%$ on the $HVV$ couplings.  

$\bullet$ The measurement of the branching ratios of the Higgs boson are of
utmost importance. For Higgs masses below $M_H \lsim 130$ GeV a large variety
of ratios can be measured at the linear collider. The $b\bar{b}, c\bar{c}$  and
$\tau^+ \tau^-$ branching ratios allow to measure the relative couplings of the
Higgs to these fermions and to check the fundamental prediction of the Higgs
mechanism that they are proportional to fermion masses. In particular BR$(H \ra
\tau^+ \tau^-) \sim m_{\tau}^2/3\bar{m}_b^2$ allows to make such a test.  In
addition, these branching ratios, if measured with enough accuracy, could allow
to distinguish a Higgs boson in the SM from its possible extensions. The
gluonic branching ratio is sensitive to the $t\bar{t}H$ Yukawa coupling [and
might therefore give an indirect measurement of this important coupling] and to
new strongly interacting particles which couple to the Higgs [such as stop in
SUSY extensions of the SM].  The branching ratio into $W$ bosons starts to be
significant for Higgs masses of the order of 120 GeV and allows to measure the
$HWW$ coupling. The branching ratio of the loop induced $\gamma \gamma$ decay
is also very important since it is sensitive to new particles [the measurement
of this ratio gives the same information as the measurement of the cross
section for Higgs boson production at $\gamma \gamma$ colliders].  

$\bullet$ The Higgs coupling to top quarks, which is the largest coupling in
the SM, is directly accessible in the process where the Higgs boson is radiated
off top quarks, $\ee \ra t\bar{t}H$. For $M_H \lsim 130$ GeV, the Yukawa
coupling  can be measured with a precision of less than 5\% at $\sqrt{s}\sim
800$ GeV with a luminosity $\int {\cal L} \sim 1$ ab$^{-1}$.  For $M_H \gsim
350$ GeV, the $Ht \bar{t}$ coupling can be derived by measuring the $H \ra
t\bar{t}$ branching ratio at higher energies.  

$\bullet$ The total width of the Higgs boson, for masses less than $\sim 200$
GeV, is so small that it cannot be resolved experimentally. However, the
measurement of BR($H \ra WW$) allows an indirect determination of $\Gamma_H$
since the $HWW$ coupling can be determined from the measurement of the Higgs
production cross section in the $WW$ fusion process [or from the measurement of
the cross section of the Higgs--strahlung process, assuming SU(2) invariance].
The accuracy of the $\Gamma_H$ measurement follows then from that of the $WW$
branching ratio. [$\Gamma_{\rm tot}$ can also be measured by using the processes
$H \leftrightarrow \gamma \gamma$]. 

$\bullet$ Finally, the measurement of the trilinear Higgs self--coupling, which
is the first non--trivial test of the Higgs potential, is accessible 
in the double Higgs production processes $\ee \ra ZHH$ [and in the $\ee \ra \nu
\bar{\nu}HH$ process at high energies]. Despite its smallness, the cross 
sections can be determined with an accuracy of the order of 20\% at a 500 GeV 
collider if a high luminosity, $\int {\cal L} \sim 1$ ab$^{-1}$, is available. 
This would allow the measurement of the trilinear Higgs self--coupling with 
an accuracy of the same order. 

An illustration of the experimental accuracies which can be achieved in the
determination of the mass, CP--nature, total decay width and the various
couplings of the Higgs boson for $M_H=120$ and 140 GeV is shown in Table 1 for
$\sqrt{s}=350$ GeV (for $M_H$ and the CP nature) and $500$ GeV (for
$\Gamma_{\rm tot}$ and all couplings except for $g_{Htt}$) and $\int {\cal
L}=500$ fb$^{-1}$ (except for $g_{Htt}$ where $\sqrt{s}=1$ TeV and $\int {\cal
L}=1$ ab$^{-1}$ are assumed). For the Higgs self--couplings, the error is only
on the determination of the cross section, leading to an error slightly
larger, $\sim 30\%$, on the coupling.  For the test of the CP nature of the
Higgs boson, $\Delta {\rm CP}$ represents the relative deviation from the
0$^{++}$ case. 

\begin{table}[htbp]
\renewcommand{\arraystretch}{1.8}
\caption{Relative accuracies (in \%) on Higgs boson mass, width and couplings 
obtained at TESLA with $\sqrt{s}=350,500$ GeV and $\int {\cal L}=500$ fb$^{-1}$ 
(except for top). }
\hskip3pc\vbox{\columnwidth=26pc
\begin{tabular}{|c|c|c|c|c|c|c|c|c|c|c|c|}\hline
$M_H$ (GeV) & $\Delta M_H$ & $\Delta {\rm CP}$ & $\Gamma_{\rm tot}$ &
$g_{HWW}$ & $g_{HZZ}$ & $g_{Htt}$ & $g_{Hbb}$ & $g_{Hcc}$ & $g_{H\tau \tau}$
& $g_{HHH}$  \\ \hline
$120$ & $\pm 0.033$ & $\pm 3.8$ & $\pm 6.1$ & $\pm 1.2$ & $\pm 1.2$ & 
$\pm 3.0$ & $\pm 2.2$ & $\pm 3.7$ & $\pm 3.3$ & $\pm 17$  \\ \hline
$140$ & $\pm 0.05$ & $-$ & $\pm 4.5$ & $\pm 2.0$ & $\pm 1.3$ & $\pm 6.1$ &
$\pm 2.2$ & $\pm 10$ & $\pm 4.8$ & $\pm 23$ 
\\ \hline
\end{tabular}
}
\end{table}
As can be seen, an $\ee$ linear collider with a high--luminosity is a very high
precision machine in the context of Higgs physics.  This precision would allow
the determination of the complete profile of the SM Higgs boson, in particular 
if its mass is smaller than $\sim 140$ GeV. It would also allow to distinguish
the SM Higgs particle from the lighter MSSM $h$ boson up to very high values
of the pseudoscalar Higgs boson mass, $M_A \sim {\cal O}(1~{\rm TeV})$. This
is exemplified in Fig.~13, where the $(g_{Hbb}, g_{HWW})$ and  $(g_{Hbb}, g_{H
\tau \tau})$ contours are shown for a Higgs boson mass $M_H=120$ GeV produced
and measured at a 500 GeV collider with $\int {\cal L}=500$ fb$^{-1}$. These
plots are obtained from a global fit which take into account the experimental 
correlation between various measurements \cite{M1}. In addition to the 
$1\sigma$ and $2\sigma$ confidence level contours for the fitted values of the 
pairs of ratios, the expected value predicted in the MSSM for a given range of 
$M_A$ is shown.

\begin{figure}[ht!]
\vspace*{-1.3cm}
\begin{center}
\begin{tabular}{c c}
\epsfig{file=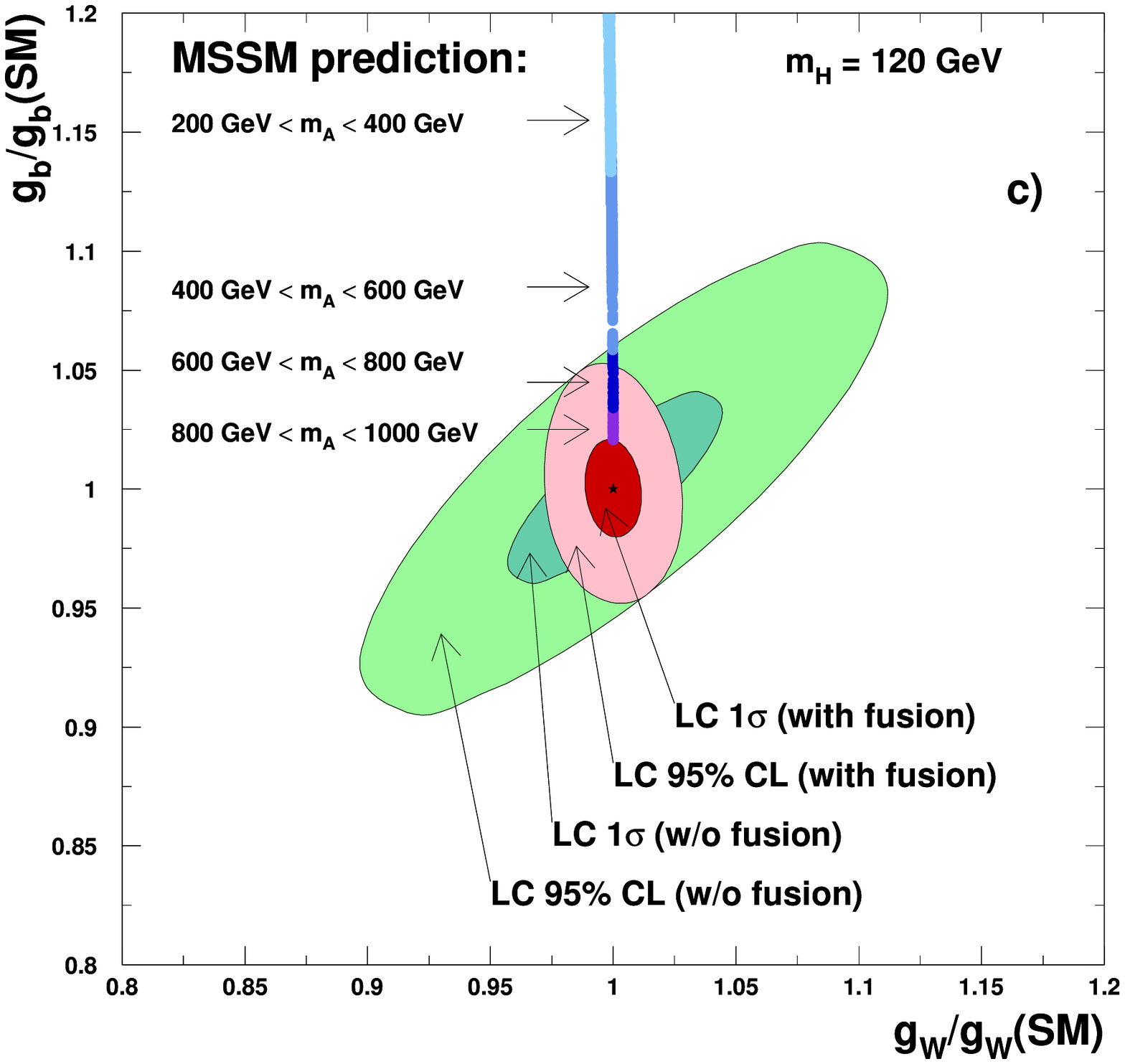,width=0.52\linewidth}\hspace*{-2mm}&\hspace*{-2mm} 
\epsfig{file=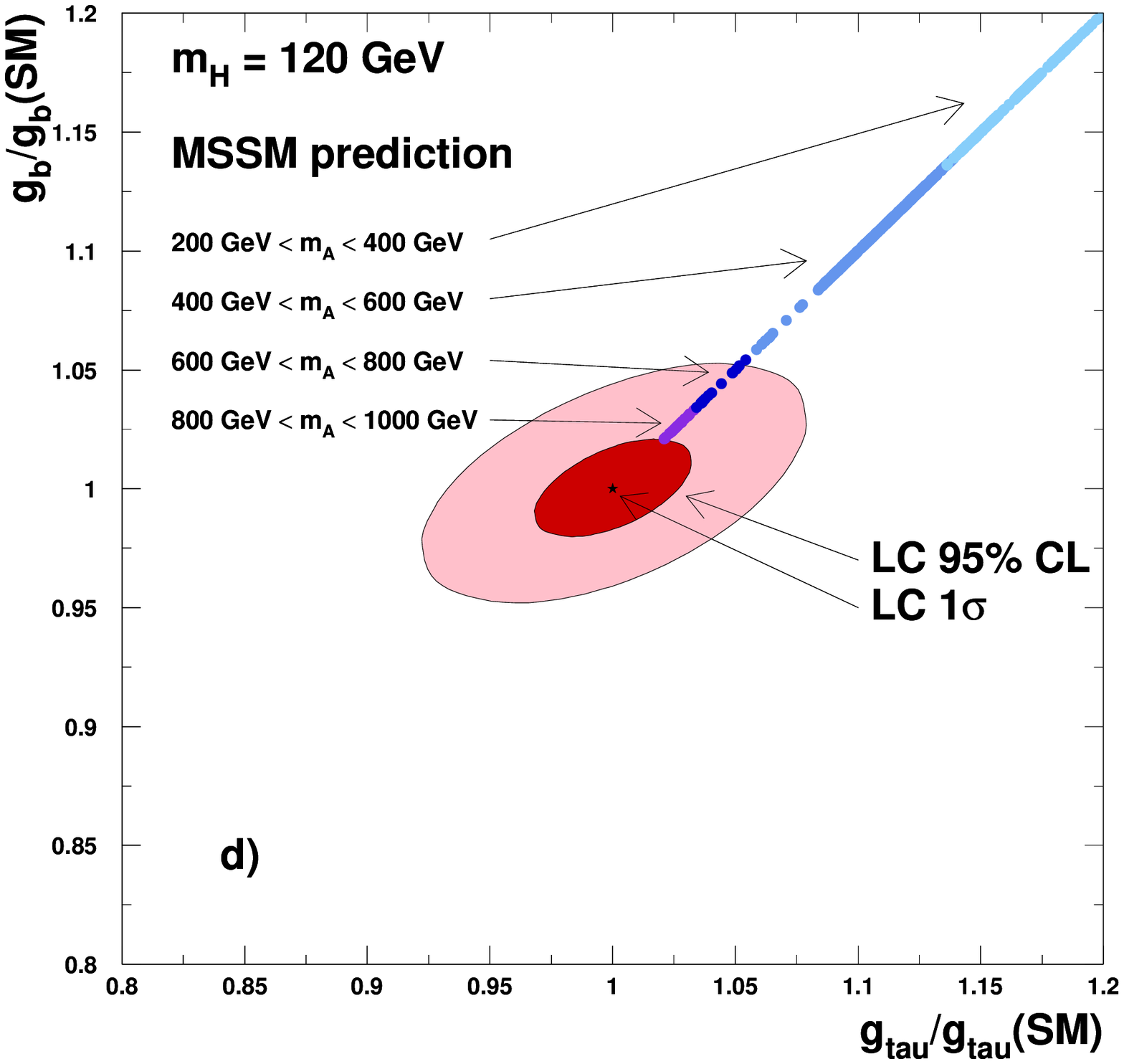,width=0.52\linewidth} \\
\end{tabular}
\caption{Higgs boson coupling determinations at TESLA for $M_H=120$ GeV
with 500~fb$^{-1}$ of data, and the expected deviations in the MSSM.}
\end{center}
\label{fig:hfitter}
\vspace*{-.3cm}
\end{figure}

\subsubsection*{5.2 Measurements at the LHC}

$\bullet$ At the LHC, the Higgs boson mass can be measured with a very good
accuracy. In the range below $M_H \lsim 400$ GeV, where the total width
is not too large, a relative precision of $\Delta M_H/M_H \sim 0.1\%$ can be
achieved in the channel $H \to ZZ^{(*)} \to 4\ell^\pm$ with 300 fb$^{-1}$
luminosity.  In the `low mass" range, a slight improvement can be obtained by
considering $H \to \gamma \gamma$. In the range $M_H \gsim 400$ GeV, the
precision starts to deteriorate because of the smaller cross sections. However
a precision of the order of $1\%$ can still be obtained for $M_H\sim 800$ GeV
if theoretical errors, such as width effects, are not taken into account.  

$\bullet$ Using the same process, $H \to ZZ^{(*)} \to 4\ell^\pm$, the total
Higgs width can be measured for masses above $M_H \gsim 200$ GeV when it is
large enough. While the precision is very poor near this mass value [a factor 
of two], it improves to reach the level of $\sim 5\%$ around $M_H \sim 400$ 
GeV. Here, again the theoretical errors are not taken into account.  

$\bullet$ The Higgs boson spin can be measured by looking at angular
correlations between the fermions in the final states in $H \to VV \to 4f$ as
in $\ee$ collisions.  However the cross sections are rather small and the
environment too difficult.  Only the measurement of the decay planes of the two
$Z$ bosons decaying into four leptons seems promising. 

$\bullet$ The direct measurement of the Higgs couplings to gauge bosons and
fermions is possible but with a rather poor accuracy. This is due to the
limited statistics, the large backgrounds and the theoretical uncertainties
from the limited precision on the parton densities and the higher order
radiative corrections or scale dependence. An example of determination of
cross sections times branching in various channels at the LHC is shown in
Fig.~14 from Ref.~\cite{R10} for a luminosity of 200 fb$^{-1}$.  Solid lines
are for $gg$ fusion, dotted lines are for $t\bar{t}H$ associated production
with $H \to b\bar{b}$ and $WW$ and dashed lines are the expectations for the
weak boson fusion process. A precision of the order of 10 to 20\% can be
achieved in some channels, while the vector boson fusion process leads to
accuracies of the order of a few percent. However, experimental analyses
accounting for the backgrounds and for the detector efficiencies as well as
further theory studies for the signal and backgrounds need to be performed to
confirm these values.  The Higgs boson self--couplings are too difficult to
measure at the LHC because of the smallness of the $gg\to HH$ and $qq \to HHZ$
cross sections \cite{M3} and the large backgrounds.  

\begin{figure}[htbp]
\vspace*{-2mm}
\hspace*{2cm} \includegraphics[width=7.6cm, angle=90]{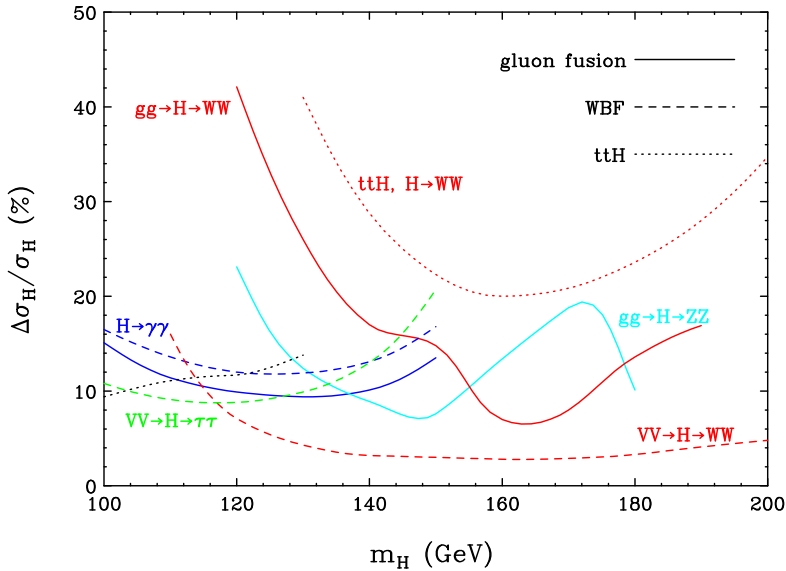}
\caption{Expected relative errors on the determination of 
$\sigma \times {\rm BR}$ for various Higgs boson search channels at the LHC 
with 200 fb$^{-1}$ data.}
 \label{fig:delsigh}
\vspace*{-.1cm}
 \end{figure}

$\bullet$ To reduce the theoretical uncertainties and some experimental errors,
it is more interesting to measure ratios of quantities where the overall 
normalization cancels out. For instance, by using the same production channel
and two decay modes, the theory error from higher order corrections and
from the poor knowledge of the parton densities drops out in the ratios. 
A few examples \cite{M2} of measurements of ratios of branching ratios or 
ratios of Higgs couplings squared at the LHC with a luminosity of 300 fb$^{-1}$
are shown in Table 2 [with still some theory errors not included in some cases].
As can be seen, a precision of the order of 10 percent can be reached in these 
preliminary analyses. 

\begin{table}[htbp]
\renewcommand{\arraystretch}{1.7}
\caption{Relative accuracies on measurements of ratios of cross sections and/or
branching ratios at the LHC with a luminosity of 300 fb$^{-1}$.}
\hskip3pc\vbox{\columnwidth=26pc
\hspace*{1.5cm}
\begin{tabular}{|c|c|l|l|}\hline
Process & Measurement quantity & Error $ \ \ \ $ &Mass range $ \ \ \ \ $ 
\\ \hline
$\frac{ (t\bar{t}H +WH) \to \gamma \gamma +X}{(t\bar{t}H +WH) \to b\bar{b}+X}$
& $\frac{ {\rm BR}(H \to \gamma \gamma)}{{\rm BR}(H \to b\bar{b})}$
& $\sim 15\%$ & $80-120$ GeV \\ \hline
$\frac{ H \to \gamma \gamma}{ H \to 4\ell^+}$
& $\frac{ {\rm BR}(H \to \gamma \gamma)}{{\rm BR}(H \to ZZ^*)}$
& $\sim 7\%$ & $120-150$ GeV \\ \hline
$ \frac{ t\bar{t}H \to \gamma \gamma, b\bar{b} }{WH\to \gamma \gamma, b\bar{b}}$
& $ \left( \frac{g_{Htt}} {g_{HWW}} \right)^2$ 
& $\sim 15\%$ & $80-120$ GeV \\ \hline
$ \frac{ H \to ZZ^* \to  4\ell^+} { H \to WW^* \to 2\ell^\pm 2\nu}$
& $ \left( \frac{g_{HZZ}} {g_{HWW}} \right)^2$ 
& $\sim 10\%$ & $130-190$ GeV \\ \hline
\end{tabular}
}
\end{table}
A more promising channel would be the vector boson fusion process, $qq\to
WW/ZZ\to Hqq$ in which $H \to \tau^+\tau^-$ or $WW^{(*)}$, which would allow
the additional measurement of the couplings to $\tau$ leptons for instance. A
preliminary parton level analysis including this production channel shows that
measurements of some Higgs boson couplings can be made at the level of 5--10\%
statistical error \cite{R10}. More work, including full detector simulation, is
needed to sharpen these analyses.  

\subsection*{6. Conclusions and Complementarity between the LHC and LC} 

In the SM, global fits of the electroweak data favor a light Higgs boson, $M_H
\lsim 200$ GeV, and if the theory is to remain valid up to the GUT scale, the
Higgs boson should be lighter than $200$ GeV. In supersymmetric extensions of
the SM, there is always one light Higgs boson with a mass $M_h \lsim 130$ GeV
in the minimal version and $M_h \lsim 200$ GeV in more general ones. Thus, a
Higgs boson is definitely accessible at next generation experiments.  

The detection of such a Higgs particle is possible at the upgraded Tevatron for
$M_H \lsim 130$ GeV and is not a problem at the LHC where even much heavier
Higgs bosons can be probed: in the SM up to $M_H \sim 1$ TeV and in the MSSM
for $M_{A,H,H^\pm}$ of order a few hundred GeV depending on $\tb$. Relatively
light Higgs bosons can also be found at future $\ee$ colliders with
c.m.\,energies $\sqrt{s} \gsim 350$ GeV; the signals are very clear and the
expected high luminosity allows to investigate thoroughly their fundamental
properties.  

In many aspects, the searches at $\ee$ colliders are complementary to those
which will be performed at the LHC.  An example can be given in the context of
the MSSM. In constrained scenarios, such as the minimal Supergravity model, the
heavier $H,A$ and $H^\pm$ bosons tend to have masses of the order of 1 TeV
\cite{M4} and therefore will escape detection at both the LHC and linear
collider. The right--handed panel of Fig.~15 shows the number of Higgs
particles in the $(M_A, \tb$) plane which can observed at the LHC and in
the white area, only the lightest $h$ boson can be observed.  In this parameter
range, the $h$ boson couplings to fermions and gauge bosons will be almost
SM--like and, because of the relatively poor accuracy of the measurements at the
LHC, it would be difficult to resolve between the SM and MSSM (or extended)
scenarii. At $\ee$ colliders such as TESLA, the Higgs couplings can be measured
with a great accuracy, allowing to distinguish between the SM and the MSSM
Higgs boson close to the decoupling limit, i.e. for pseudoscalar boson masses
which are not accessible at the LHC.  This is exemplified in the right-panel of
Fig.~15, where the accuracy in the determination of the Higgs couplings to
$t\bar{t}$ and $WW$ are displayed, together with the predicted values in the
MSSM for different values of $M_A$.  The two scenarii can be distinguished for
pseudoscalar Higgs masses up to 1 TeV.  

\begin{figure}[htbp]
\vspace*{-1.2cm}
\begin{center}
\epsfig{file=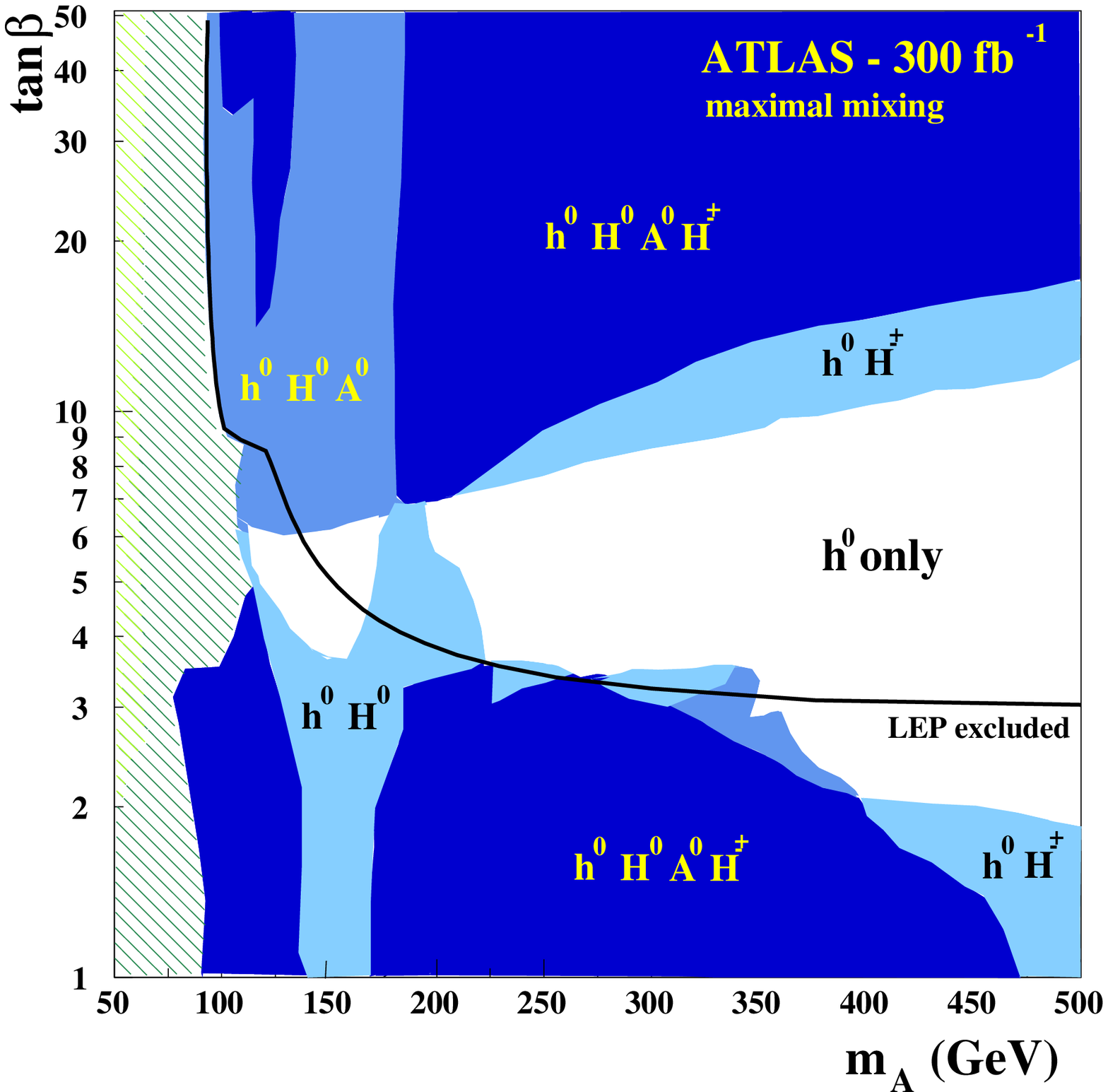,width=0.48\linewidth}\hspace*{-.5mm}
\epsfig{file=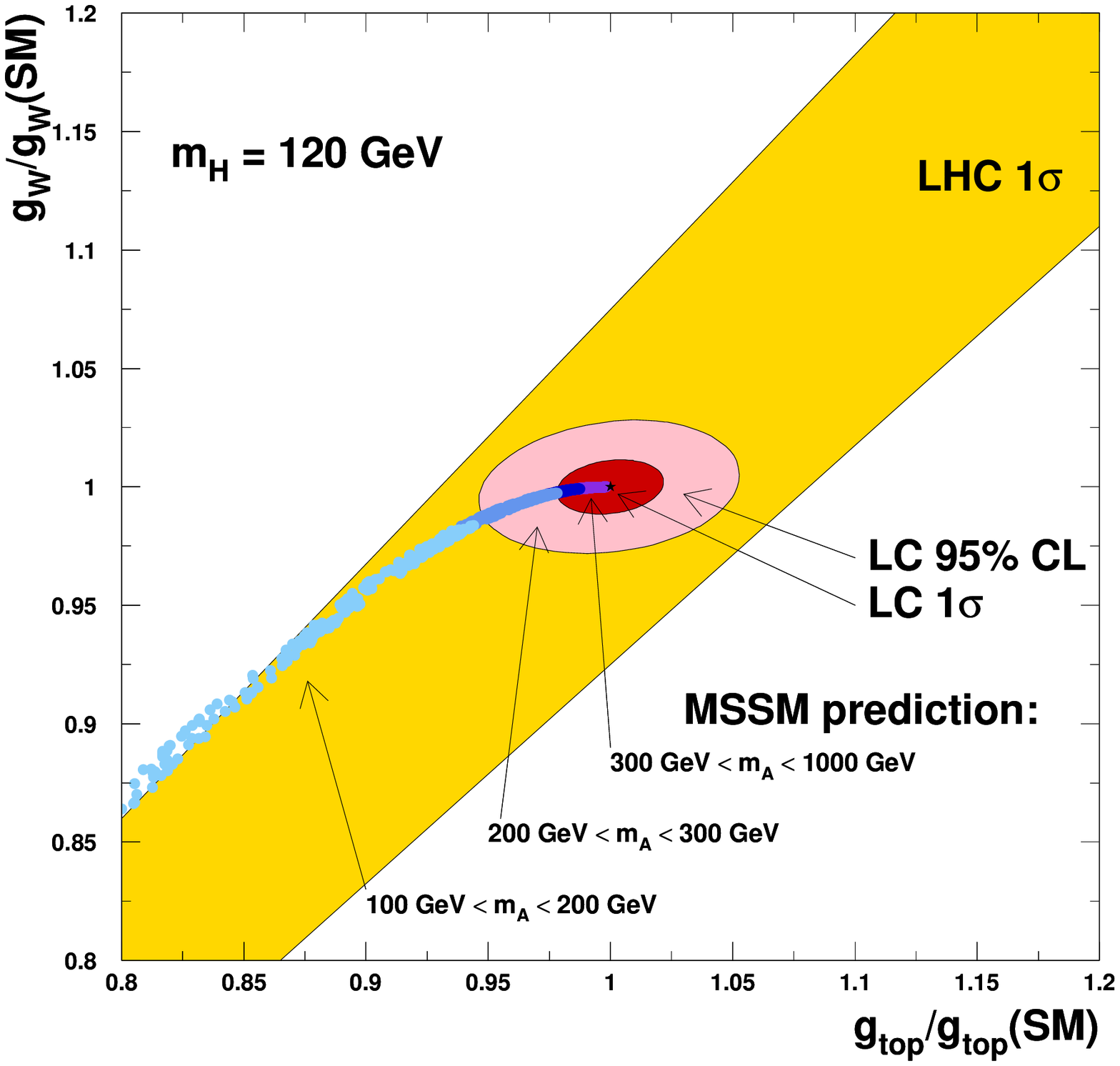,width=0.53\linewidth}
\caption{ Number of Higgs bosons which can be observed at the LHC in the
$(M_A, \tb)$ plane (right) and a comparison of the accuracy in the 
determination of the $g_{ttH}$ and $g_{WWH}$ couplings at the LHC and at TESLA 
compared to the predictions from MSSM for different values of $M_A$.}
\end{center}
\label{fig:lhclcgh}
\vspace*{-.1cm}
\end{figure}

\nn {\bf Acknowledgments}: I thank the organizers of the WHEPP VII workshop in
Allahabad, in particular Biswarup Mukhopadhyaya and Rohini Godbole, for the
invitation to the meeting and for the very nice and warm atmosphere. I thank
also the people who attended the workshop for very nice (in particular after
dinner) physics and non--physics discussions. Special thanks go to the 
``guru" K. Sridhar for taking care of our souls and our stomachs.

\end{document}